%% file: paper.tex
\documentclass{ar-1col}

\newcounter{defcounter}
\usepackage{amsthm}

\newenvironment{tbequation}{%
\addtocounter{equation}{-1}
\refstepcounter{defcounter}
\renewcommand\theequation{SB\thedefcounter}
\begin{equation}}
{\end{equation}}
\usepackage[comma]{natbib}
\usepackage{url,bm}
\usepackage[utf8]{inputenc}
\usepackage{fourier} 
\usepackage{array}
\usepackage{makecell}
\usepackage[makeroom]{cancel}
\usepackage{amssymb}
\usepackage[flushleft]{threeparttable}

\graphicspath{{./fig/}{./png/}}

\input macros

\jname{\today}
\jvol{AA}
\jyear{YYYY}
\doi{10.1146/((please add article doi))}

\begin{document}
\markboth{Brandenburg \& Ntormousi}{Galactic Dynamos}

\title{Galactic Dynamos}

\author{Axel Brandenburg$^{1,2,3,4}$ \& Evangelia Ntormousi$^{5,6}$
\affil{$^1$Nordita, KTH Royal Institute of Technology and Stockholm University,
Hannes Alfv\'ens v\"ag 12, SE-10691 Stockholm, Sweden}
\affil{$^2$The Oskar Klein Centre, Department of Astronomy,
Stockholm University, AlbaNova, SE-10691 Stockholm, Sweden}
\affil{$^3$McWilliams Center for Cosmology \& Department of Physics,
Carnegie Mellon University, Pittsburgh, PA 15213, USA}
\affil{$^4$School of Natural Sciences and Medicine, Ilia State University,
3-5 Cholokashvili Avenue, 0194 Tbilisi, Georgia}
\affil{$^5$Scuola Normale Superiore, Piazza dei Cavalieri 7, I-56126 Pisa, Italy}
\affil{$^6$Institute of Astrophysics, Foundation for Research and Technology-Hellas, Vasilika Vouton, GR-70013 Heraklion, Greece}
}

\begin{abstract}
Spiral galaxies, including the Milky Way, have large-scale magnetic
fields with significant energy densities.
The dominant theory attributes these magnetic fields to a large-scale dynamo.
We review the current status of dynamo theory and discuss various
numerical simulations designed to explain either particular aspects
of the problem or to reproduce galactic magnetic fields globally.
Our main conclusions can be summarized as follows.
\parbox{10cm}{
\begin{itemize}
\item 
Idealized direct numerical simulations produce mean magnetic fields,
whose saturation energy density tends to decline with
increasing magnetic Reynolds number.
This is still an unsolved problem.
\item 
Large-scale galactic magnetic fields of microgauss strengths
can probably only be explained if helical magnetic fields of small or
moderate length scales can rapidly be ejected or destroyed.
\item
Small-scale dynamos are important throughout a galaxy's life, and probably
provide strong seed fields at early stages.
\item 
The circumgalactic medium (CGM) may play an important role
in driving dynamo action at small and large length scales.
These interactions between the galactic disk and the CGM
may provide important insights into our understanding of galactic dynamos.
\end{itemize}}
We expect future research in galactic dynamos to focus on the cosmological
history of galaxies and the interaction with the CGM as
means of replacing the idealized boundary conditions used in earlier work.
\end{abstract}

\maketitle

\tableofcontents

\section{INTRODUCTION}

Many spiral galaxies have microgauss magnetic fields, so that their
magnetic energy densities are comparable to the thermal, kinetic, and
cosmic ray energy densities; see \cite{RSS88} for an early book on
the subject, and \cite{SS22} for a recent one; hereafter SS22.
Similar magnetic field strengths have also been detected in galaxies at
larger redshifts up to $z\simeq1$ \citep{Bernet_2008,Mao2017}.

Galactic magnetic fields often also show large-scale coherence.
The first evidence for a global Galactic magnetic field comes from
optical polarization \citep{Hiltner49, Hall49}.
The existence of magnetic fields for other galaxies was later confirmed
using synchrotron emission \citep{Segalovitz+76}, which showed systematic
large-scale magnetic fields roughly in the direction of the galactic
spiral arms.
There has long been a debate about the origin of such magnetic fields: are
they primordial or dynamo-generated, or perhaps a combination of the two?
\begin{marginnote}[]
\entry{Dynamos}{convert kinetic energy into magnetic energy}
\end{marginnote}
Over the past few decades, attention has shifted from a
primordial to a dynamo-generated origin.
In the meantime, however, there have also been significant developments
in dynamo theory, and global numerical simulations are now becoming
more realistic.
They tend to show that large-scale magnetic fields can be generated by
a dynamo, but the amplitudes may be insufficient or the timescales for
their generation too long for the simulations presented so far.

In this review, we focus on galactic dynamos and highlight
the main developments since the time of the review of \cite{Beck+96}.
The broader problem of galactic magnetism that was addressed there will
not be reviewed; we refer readers to the reviews by \cite{Beck01,Beck15},
\cite{Beck+Wielebinski13}, and \cite{Han17}, and the book by SS22.
A review of astrophysical dynamos covering the era before 2005 is
given by \cite{BS05}.
The mathematics of small-scale turbulent dynamos is explained in the
book by \cite{ZRS90}, and those in partially ionized plasmas are
discussed by \cite{XL21}.
We also recommend the reviews on ISM magnetic fields by \citet{Crutcher2012},
\citet{Hennebelle_Inutsuka2019}, and \citet{pattle2022}.

At the time of the review by \cite{Beck+96},
there were results suggesting that the dynamo effect in
mean-field theory is ``catastrophically'' quenched, i.e., it goes
to zero as the magnetic Reynolds number ($\Rm$) becomes large \citep{VC92,CH96}.
\begin{marginnote}[]
\entry{Catastrophic quenching means}{$\alpha\to0$ as $\Rm\to\infty$,
so it limits the large-scale field at large $\Rm$}
\end{marginnote}
Specifically, the mean-field effect in question has been termed
the $\alpha$ effect, which quantifies the component of the mean
electromagnetic force in the direction of the mean magnetic field.
More generally, however, it means that the resulting mean (or large-scale)
magnetic field cannot be generated at the expected amplitudes or time
scales.
This led to a major crisis in dynamo theory, questioning the
possibility of an $\alpha$ effect dynamo in the nonlinear regime beyond
just infinitesimally weak kinematic dynamo-generated magnetic fields.

Although there are still unresolved questions in nonlinear dynamo
theory today, there have also been major developments in this field: the
importance of magnetic helicity fluxes has been recognized, mean-field
dynamo coefficients can now be determined from simulations without the
restrictions imposed by analytic techniques, and new dynamo mechanisms
beyond just the $\alpha$ effect have been explored.
\begin{marginnote}[]
\entry{Magnetic helicity}{Volume-integrated dot product of magnetic
vector potential $\AAA$ and magnetic field $\BB$, $\int\AAA\cdot\BB dV$}
\end{marginnote}
At the same time, there has been significant progress in performing
realistic three-dimensional (3-D) magnetohydrodynamic (MHD) simulations of
galaxy formation, allowing a new theoretical view of the problem, where
the circumgalactic medium (CGM) plays an integral part.
All these developments motivate a new review on galactic dynamos.

\begin{textbox}[t]\section{Observational tracers of galactic magnetic fields}
\paragraph*{Dust extinction / emission polarization.}

Elongated dust grains in the interstellar medium tend to align their minor
axes with the mean magnetic field direction
\citep{davis_Greenstein_1951}.
As a result, the light they emit in infrared wavelengths is polarized,
with the polarization direction perpendicular to the mean direction of
the magnetic field.
Only recently has it been possible to trace extragalactic magnetic fields
with this method \citep[e.g.,the recent survey by][]{LR2022}.

Since starlight emission is unpolarized, we can also measure dust
polarization in absorption against stellar sources.
If the distances to the stars are known, it is possible to map
the magnetic field at different locations along the line-of-sight
\citep[e.g., the Polar-Areas Stellar-Imaging in Polarization High-Accuracy Experiment (PASIPHAE)  survey,][]{Tassis2018}.

\paragraph*{Synchrotron emission.}

Ultra-relativistic cosmic ray particles emit polarized synchrotron
radiation in radio wavelengths as they gyrate around the galactic
magnetic field.
Synchrotron emission, which yields the plane-of-the-sky magnetic field
component in the warm/hot ISM, has been the tracer of choice for studying
extragalactic magnetic fields \citep[see, e.g.,][for a review]{Beck_2012}.

In general, linear polarization is measured through the Stokes parameters
$I$, $Q$, and $U$.
Then the observed polarization angle $\chi$ is calculated from the
expression $\chi = \frac{1}{2}\arctan(U/Q)$ and the polarization
fraction is $p=\sqrt{Q^2+U^2}/I$.

\paragraph*{Faraday Rotation.}

When polarized radiation passes through an ionized magnetized medium,
the plane of polarization is rotated by an angle $\Delta\chi = RM\lambda^2$,
where $\lambda$ is the observed wavelength and
$RM=0.81 \int_0^D n_{\rm e} B_{\parallel} \dd l$
is in units of radians/$m^2$.
Here, $D$ is the distance to the source in pc,
$n_{\rm e}$ is the electron density in $\cm^{-3}$, and $B_{\parallel}$ is
the line-of-sight magnetic field of the medium in 
microgauss
\citep{Burn_1966, Brentjens_deBruyn2005}.
If we have an estimate of the electron density, then Faraday
rotation can yield the line-of-sight magnetic field.

\end{textbox}

\section{DYNAMOS}

We begin by discussing historical and theoretical aspects of MHD and
dynamos that are of particular importance in connection with the new
developments in galactic dynamo research over the past few decades.
The presence of a microphysical magnetic diffusivity plays here an
important role.
We begin with a historical perspective.

\subsection{Historical remarks}
\label{sec:history}

In \Sec{sec:mean-field_models} we will discuss the observational
signatures of dynamo models in detail.
Here, we will only briefly mention some observational results that
were significant for the development of dynamo models.
For later reference, we provide here a short overview of the existing
galactic magnetic field tracers in the text box on ``Observational
tracers of galactic magnetic fields''.

\begin{figure}[t]
\includegraphics[width=\textwidth]{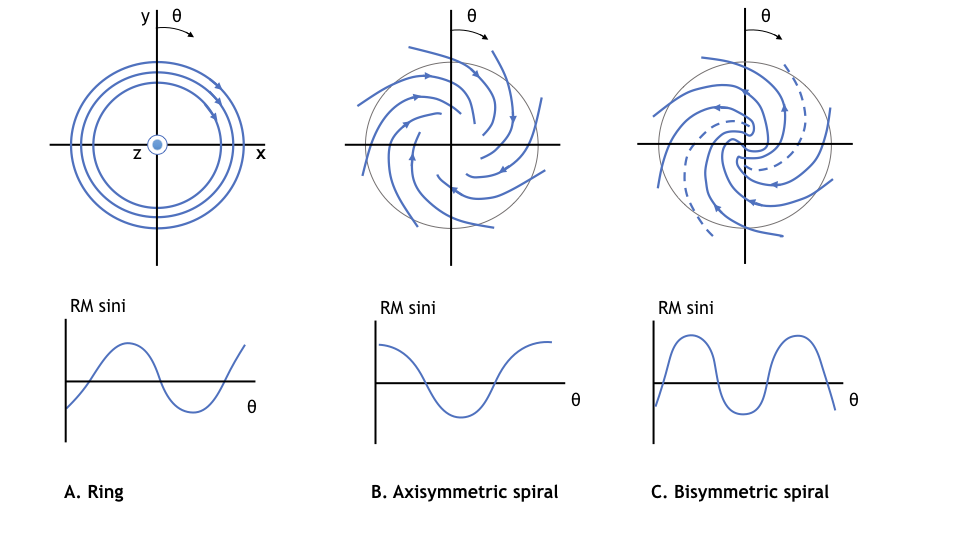}
\caption{
Sketch of the rotation measure (RM) signature of a tilted galaxy with
a ring, axisymmetric, or bisymmetric magnetic field.
The inclination $i$ is the angle between the $z$ axis, indicated in
the top left panel, and the line of sight.
Only when $i\neq0$ can one see the RM signature as sketched in the
bottom panels.
}\label{sketch_aas_bss}
\end{figure}

In the late 1970s, synchrotron radiation from external spiral galaxies
began to reveal the presence of large-scale ordered magnetic fields
broadly aligned with the galactic spiral pattern.
At the time, an obvious possibility was that these magnetic fields were
the result of winding up a pre-existing, large-scale field. 
This idea goes back to \cite{Pid64} and \cite{Oki+64}, and is now
commonly termed the primordial origin of the magnetic field.
The resulting magnetic field is then expected to have the form of a
bisymmetric spiral (BSS).
The BSS form has a characteristic signature in the Faraday Rotation
Measure (RM): when we observe an external galaxy almost, but not exactly,
face-on, the line-of-sight magnetic field measured through RM probes
the azimuthal component of the galactic field -- see \Fig{sketch_aas_bss}
for a sketch.
Using this technique, \citet{TF78}, who used a sketch similar to
ours in \Fig{sketch_aas_bss}, found evidence for a BSS in M51.
In an early review on galactic magnetic fields, \cite{SFW86} contrasted
BSS with an axisymmetric spiral (ASS), expected from mean-field dynamo
theory.
\Fig{sketch_aas_bss} also sketches the expected ASS signature.

\begin{figure}[t]
\includegraphics[width=\textwidth]{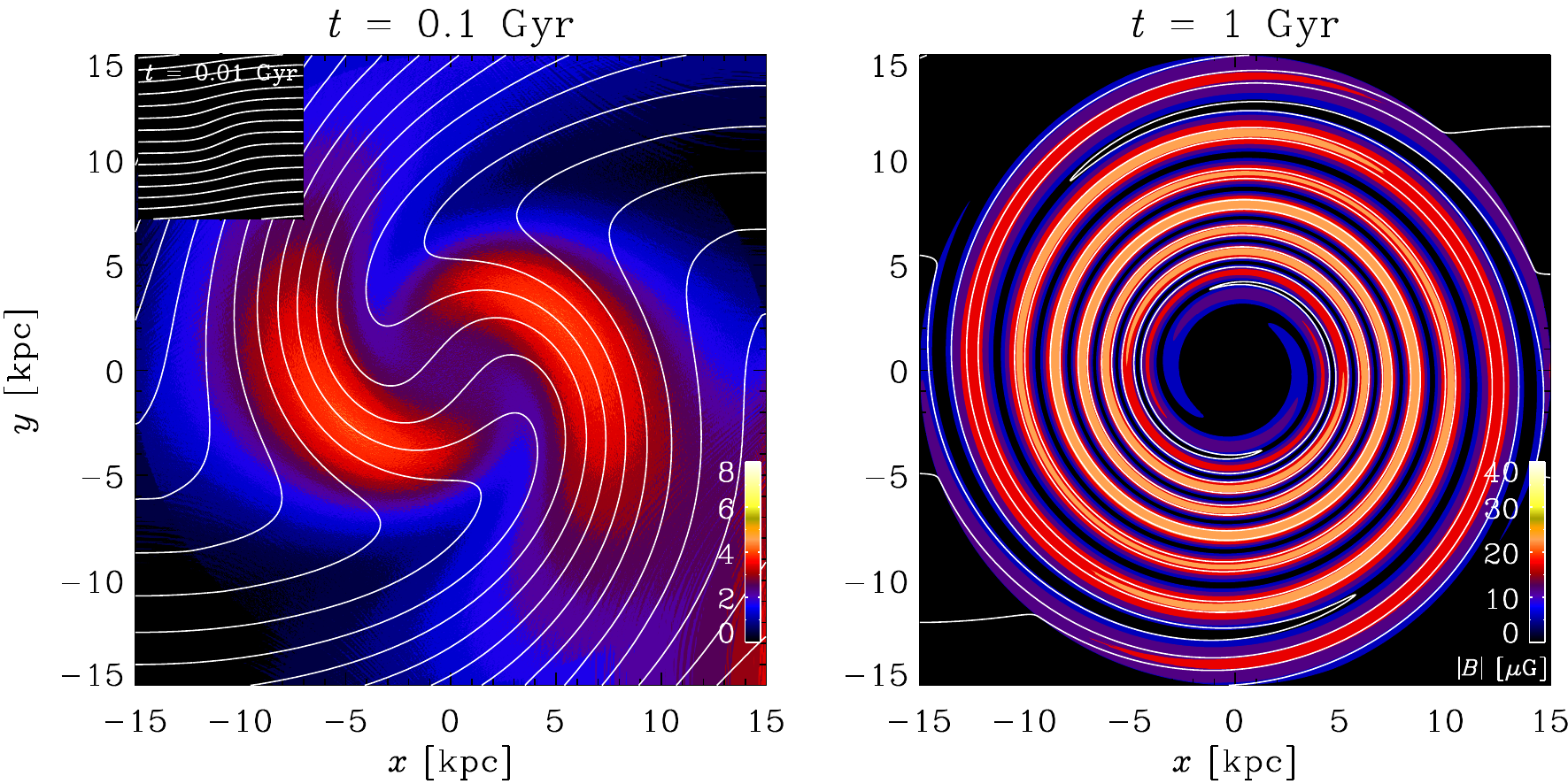}
\caption{
Snapshots of field lines together with representations of $|\BB|$
color-coded (in units of its original value) at $0.1$ and $1\Gyr$
for the wind-up problem described in the text.
The inset in the upper left corner of the left panel shows the field lines at the time
$0.01\Gyr$.
}\label{psavss}
\end{figure}

However, a purely primordial origin of galactic magnetic fields implies
that a tremendous amount of winding has occurred over the past $14\Gyr$
due to the shear induced by the differential rotation of the galaxy.
For example, in the solar neighborhood, the angular velocity of the
Galaxy is $\approx30\Gyr^{-1}$, i.e., the rotation period is
$(2\pi/30)\Gyr\approx0.2\Gyr$.
This yields 70 revolutions in $14\Gyr$, so we would expect the magnetic
field to be strongly wound up.
\FFig{psavss} gives a quantitative illustration of this wind-up process.
It shows color scale images of $|\BB|$ together with field lines
corresponding to the contours of the normal component of the magnetic
vector potential, $A_z(x,y)$, so that the magnetic field in the plane
is given by $\BB=\nab\times(\zzz A_z)$.
To obtain the result shown in \Fig{psavss}, we solved the two-dimensional
(2-D) induction equation, which corresponds to an advection--diffusion
equation of the form
\begin{equation}
\frac{\DD A_z}{\DD t}=\eta\nabla^2 A_z,
\label{InductEq2D}
\end{equation}
where $\DD/\DD t=\partial/\partial t+\UU\cdot\nab$. 
Here, we assumed that $\UU=\Omega(x,y)\pom$, where
$\pom=(x,y,0)$ is the cylindrical position vector and
$\Omega=\Omega_0/[1+(\varpi/\varpi_0)^{n}]^{1/n}$ is the angular velocity
with $\varpi_0=5\kpc$, $n=3$, and $\Omega_0=40\Gyr^{-1}$.
This experiment demonstrates the extreme winding of the magnetic field.
The turbulent magnetic diffusivity is here $5\times10^{-4}\kpc\,\km\s^{-1}$,
corresponding to $1.5\times10^{23}\cm^2\s^{-1}$, which is three orders
of magnitude below the canonical estimates \citep[][SS22]{BDMSST93}.
We see about six windings at $1\Gyr$ with a thirty-fold increase of $|\BB|$.
This amount of winding is not observed in any galaxy.

Mean-field dynamo theory was originally developed in the solar context
\citep{SK69} and can predict both axisymmetric and nonaxisymmetric
magnetic fields \citep{Baryshnikova+87}.
\cite{Par71} was the first to show that the most easily excited
axisymmetric large-scale magnetic fields in oblate bodies such as galaxies
have an azimuthal component that is symmetric about the midplane, i.e.,
the fields have quadrupolar symmetry -- in contrast to the dipolar
symmetry that is often found in spherical bodies such as the Earth.

\begin{figure}\centering\includegraphics[width=\textwidth]{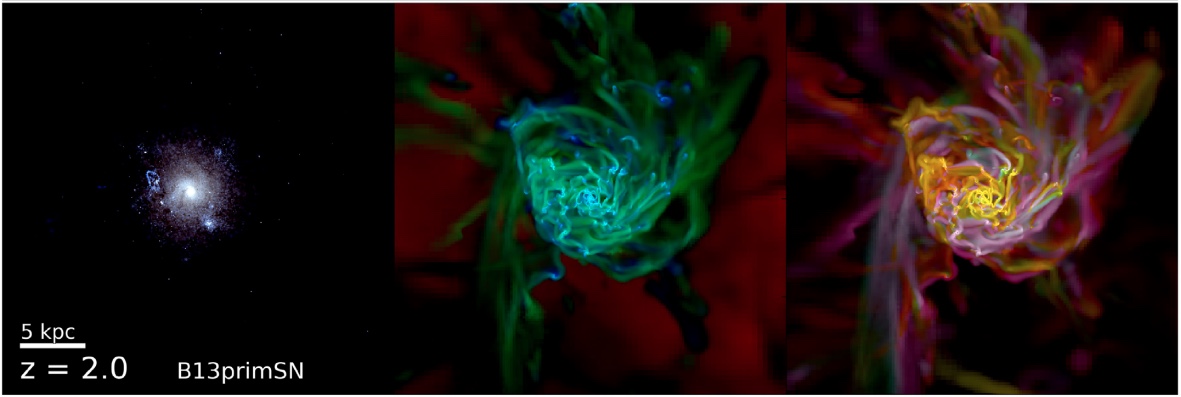}
\caption{Section of Figure~1 from \citet{MartinAlvarez2021}, showing a cosmological
galaxy model evolved with two different initial magnetic fields:primordial
or ``injected'' on small scales by stellar feedback.
The left panel shows a color-composite mock observation in the optical,
the middle panel shows dust absorption along the line of sight, and the
one on the right color-codes the total magnetic energy according to its
origin: green for the primordial, red for the injected, and blue for
the cross-term field.}
\label{fig:martinalvarez}
\end{figure}

While dynamo theory can also produce BSS-type fields
\citep{Krasheninnikova+89}, they are not the most
easily excited modes \citep{Elstner+90, BTK90}, unless the turbulence is
strongly anisotropic or the dynamo is controlled by strongly anisotropic
flow structures \citep{Moss+93}.
Today, the significance of primordial magnetic fields is still
not resolved, and global 3-D numerical simulations
suggest that both primordial and magnetic fields of astrophysical
origin may be present in typical galaxies \citep[e.g.,][see
\Fig{fig:martinalvarez}]{MartinAlvarez2021}.

\subsection{The need for magnetic diffusivity: the example of steady flows}
\label{NeedMagneticDiffusivity}

The evolution of the magnetic field $\BB$ is governed by the
usual induction equation,
\begin{equation}
\frac{\partial\BB}{\partial t}=\nab\times\left(\UU\times\BB-\JJ/\sigma\right),
\label{InductEq}
\end{equation}
where $\UU$ is the velocity and $\JJ=\nab\times\BB/\mu_0$ is the current
density with $\mu_0$ being the vacuum permeability.
\EEq{InductEq} also includes an electric conductivity $\sigma$, because
the mean-free path of the electrons in the interstellar medium (of the
order of a few thousand astronomical units) is much smaller that the
scales under study here.
The magnetic resistivity is $1/\sigma$, and the microphysical
magnetic diffusivity is then given by $\eta=1/\sigma\mu_0$.

\begin{textbox}[t]\section{Characteristic nondimensional numbers}
Fluid and magnetic Reynolds numbers and their ratio, the magnetic
Prandtl number, are defined as
\begin{tbequation}
\Rey=\urms/\nu\kf,\quad
\Rm=\urms/\eta\kf,\quad
\Pm=\nu/\eta,
\label{Numbers}
\end{tbequation}
where $\urms$ is the root-mean square velocity,
$\nu$ is the (microphysical) kinematic viscosity,
$\eta$ is the (microphysical) magnetic diffusivity,
and $\kf$ is the characteristic flow wavenumber.
Note that the Reynolds numbers are sometimes based on the length scale,
$2\pi/\kf$, which leads to about six times larger values.
The present definition is commonly used in numerical simulations of
turbulence.
As a rule of thumb, the number of mesh points needed in a numerical
simulation is similar to the value of the Reynolds number.
In simulations with partial ionization, the ionization ratio
enters as another nondimensional number.
\end{textbox}

Dynamos convert kinetic energy into magnetic energy through what is
termed the dynamo instability.
It occurs when the magnetic Reynolds number, $\Rm$,
exceeds a certain critical value (for the definition of $\Rm$, see the
text box titled ``Characteristic nondimensional numbers'').
Here, $\kf$ is the typical wavenumber of the flow.
A rigorous definition of this instability is only possible for steady flows.
Then an eigenvalue problem can be expressed through:
$\BB(\xx,t)=\BB_\lambda(\xx)\,e^{\lambda t}$, where $\lambda$ is the
eigenvalue and $\BB_\lambda(\xx)$ the eigenfunction.
For steady, mass conserving, compressible flows, \cite{MP85} proved
that dynamos (i.e., $\lambda>0$) cannot exist for $\eta=0$, i.e., in
the strictly ideal case.
This does not preclude dynamos in the astrophysically relevant limit
$\eta\to0$, which are called fast dynamos \citep{Sow87}, but it is
important to stress that the {\em limit} $\eta\to0$ is quite different
from the {\em case} $\eta=0$.

The case $\eta=0$ is arguably pathological, 
because without resistivity, there is no Joule heating
and the field line topology cannot change.
This case is therefore of academic interest only,
although it can be described using
Euler potentials, $\Phi(\xx,t)$ and $\Psi(\xx,t)$, such that
$\BB(\xx,t)=\nab\Phi\times\nab\Psi$, where $\Phi(\xx,t)$ and
$\Psi(\xx,t)$ obey \citep[e.g.,][]{RP07},
\begin{equation}
\frac{\DD\Phi}{\DD t}=
\frac{\DD\Psi}{\DD t}=0\quad
\Longleftrightarrow\quad
\frac{\partial\BB}{\partial t}=\nab\times\left(\UU\times\BB\right).
\end{equation}
We see that in the special case of 2-D, this equation agrees with
the advection--diffusion equation \Eq{InductEq2D}, where $\Phi=A_z$
and $\Psi=z$ have been assumed.
In this special case, it is possible to recover the induction equation
in the presence of microphysical magnetic diffusion.

Dynamos have {\em not} been found in this formulation -- even for
3-D turbulent or other flows that allow for dynamo action
in the limit $\eta\to0$ \citep{Bra10}.
The method forbids even a very weakly diffusive advection of $\Phi$
and $\Psi$, which would be needed in any numerical simulation to prevent
the formation of infinitely sharp gradients.

To understand the problem with the case $\eta=0$, let us now discuss
instead the limit $\eta\to0$.
The tangling of a pre-existing magnetic field {\em can}
convert kinetic energy into magnetic energy for some period of time, but
it is then not through a dynamo instability, and can happen for a purely
2-D field, $\BB=\BB(x,y)$, as we have seen in \Fig{psavss}.
The magnetic field is amplified by perpetual stretching, so it
continuously develops smaller scale structures. 
This continuous change in the field structure makes it impossible to 
describe the evolving field
by an eigenfunction of the form $\BB_\lambda(\xx)$, which is
independent of time.
The actual solution $\BB(\xx,t)$ in the case of $\eta=0$ would
continuously develop smaller length scales as time goes on.
Thus, even though the growth may still be exponential, the solution
cannot be separated into a purely temporal and a purely spatial part.

\subsection{Dynamos in turbulent and time-dependent flows}

All astrophysically relevant flows are time-dependent.
Turbulent flows can be statistically steady, so one can still determine
an eigenvalue problem by averaging over the fluctuations; see \cite{SB14}
for detailed studies of kinematic dynamos in helical and fractionally
helical turbulence at large magnetic Reynolds numbers.
Even in those turbulent time-dependent flows, when eigenvalues and
statistical eigenfunctions with certain energy spectra are obtained
empirically for finite $\eta$ by suitable averaging, no dynamos have
been found in the case $\eta=0$, when Euler potentials can be used,
as discussed above.

In practice, we are often also interested in decaying or collapsing
turbulent flows.
Dynamos may occur in those cases, but they are hard to define
rigorously.
Nevertheless, amplification---suggestive of dynamo action---both for
decaying turbulence \citep{Bran+19} and for turbulent gravitational
collapse \citep{Sur+12, XL20} has been reported, as is discussed in
\Sec{sec:grav_collapse} of this review.

\begin{textbox}[t]\section{The $\alpha$ effect: example of a mean-field dynamo}
The $\alpha$ effect quantifies how a systematic twist (or swirl) in a
turbulent flow produces secondary magnetic fields around a primary field
in a specific direction.
An example is the production of a poloidal field from a toroidal field,
as is believed to occur through cyclonic convection in the Sun \citep{Par55}.
Mathematically, this is described by a contribution to the
averaged electromotive force, $\overline{\uu\times\bb}$,
in the direction of the main magnetic field $\meanBB$, i.e.,
$\overline{\uu\times\bb}=\alpha\meanBB+$ higher-order derivatives.
It is called $\alpha$ effect, because of the historically chosen
coefficient $\alpha$.
Here, $\uu$ and $\bb$ are fluctuations of $\UU$ and $\BB$, respectively.
The type of averaging depends on the problem at hand and will be
discussed later in \Sec{Averaging}.
Using $\meanBB=\nab\times\meanAA$, and ignoring for now mean
flows such as the galactic differential rotation, the averaged
uncurled induction equation \eq{InductEq} takes the form
\begin{tbequation}
\partial\meanAA/\partial t=\alpha\nab\times\meanAA+\eta\nabla^2\meanAA,
\label{dAmeandt}
\end{tbequation}
where $\eta=\const$ has been assumed.
For $\alpha=\const$, solutions are proportional to the eigenfunctions
of the curl operator, for example $\meanAA=(\sin kz,\,\cos kz,\,0)$,
which satisfies $\nab\times\meanAA=k\meanAA$.
Seeking solutions of the form $\meanAA\propto\AAA_0 e^{\ii kz+\lambda t}$,
with $\AAA_0$ being the eigenfunction, yields the dispersion relation
$\lambda=\alpha k-\eta k^2$, and therefore self-excited solutions for
$\alpha>\eta k$.

The $\alpha$ effect thus explains the exponential growth of a weak
mean magnetic field.
We recall that the full magnetic field has fluctuations, but they are
usually growing at the same rate as the mean field.
Since the magnetic field is a pseudovector, but the electromotive force
is an ordinary vector, $\alpha$ must be a pseudoscalar, i.e., its sign
changes when viewed in a mirror.
An $\alpha$ effect can occur when the system is governed by a specific
pseudoscalar \citep{KR80}.
As an example, systems governed by gravity $\grav$ and angular velocity
$\OO$ have a finite pseudoscalar given by $\grav\cdot\OO$.
The existence of this pseudoscalar is is what caused the systematic
twist or swirl in the flow which, in turn, produces the $\alpha$ effect
in galaxies.
Twist or swirl can also occur through corresponding driving and through
initial conditions.
It is then characterized by the kinetic helicity.

One of the higher-order derivative contributions to
$\overline{\uu\times\bb}$ is from turbulent diffusion, so one has
$\overline{\uu\times\bb}=\alpha\meanBB-\etat\mu_0\meanJJ$, where
$\etat$ is the turbulent magnetic diffusivity.
The second term nearly balances the former and is therefore important.
We also note that two further generalizations to this formulation will
be discussed in \Sec{ParameterizationEMF}: (i) $\alpha$ and $\etat$ become
tensors and (ii) the multiplications become convolutions.
\end{textbox}

\subsection{Early examples of dynamos}
\label{SimpleDynamos}

\citet{Cow33} formulated an anti-dynamo theorem, concluding that
``The theory proposed by Sir Joseph Larmor, that
the magnetic field of a sunspot is maintained by the currents it induces
in moving matter, is examined and shown to be faulty.''
At that time, there was no
hint that the solution to the problem could lie in the third dimension.
It was only later that the use of a 2-D analysis in the work
of \cite{Cow33} was understood as not just a simplification, but as a
crucial restriction precluding dynamo action.
Even after Parker's discovery \citep{Par55} of what is now called the
$\alpha$ effect (see the text box on ``The $\alpha$ effect''),
it was not generally accepted that dynamos could work even in principle.
For example, \cite{Cha56} found that particular flow geometries could
prolong the resistive decay time to half a billion years when using the
magnetic diffusivity of the Earth's outer core.
He speculated that the Earth's magnetic field could be explained in
that way, rather than by a dynamo.
His speculation suggests that the existence of dynamos was far from
being widely accepted at that time.

\begin{marginnote}[]
\entry{Small-scale dynamos}{generate magnetic fields at the resistive scale
in the kinematic regime, but at a fraction of the forcing scale otherwise}
\end{marginnote}

\begin{marginnote}[]
\entry{Large-scale dynamos}{create coherent structures on large spatial
and temporal scales}
\end{marginnote}

The first rigorous examples of dynamos were presented by
\cite{Herzenberg58} and \cite{Backus58}.
The former, consisting of two rotors (eddies) with an angle between
their axes, was also realized experimentally \citep{Lowes+Wilkinson63}.
However, the length scale of those magnetic fields was only comparable
to that of the rotors.
This property could classify the Herzenberg result as a small-scale dynamo.
During the kinematic phase, small-scale dynamos produce a field at the
resistive scale and can later grow to the scale of turbulent eddies as
the dynamo saturates.
They do not possess a mean field.

In the early 1970s, \citet{Rob72} showed that several non-planar 2-D,
spatially periodic steady flows can exhibit dynamo action.
These flows are now called Roberts flows~I--IV.
They are large-scale dynamos and their properties have been investigated
with modern tools \citep{RDRB14}.
The expressions for the four Roberts flows are included in a dedicated
text box.
Flow~I has maximum kinetic helicity with $\bra{\oo\cdot\uu}=\kf\bra{\uu^2}$,
where angle brackets denote volume averaging.
Flow~II has $\oo\cdot\uu=0$ pointwise, while flows~III and IV
have vanishing helicity only on average ($\bra{\oo\cdot\uu}=0$), but not pointwise.
\begin{marginnote}[]
\entry{Kinetic helicity}{$\bra{\oo\cdot\uu}$, where $\oo=\nab\times\uu$
is the vorticity}
\end{marginnote}

\begin{textbox}[t]\section{The four Roberts flows}
The four Roberts flows are classic examples of large-scale dynamos.
They serve as simple benchmarks and highlight the existence of completely
different mechanisms.
Only the first one corresponds to the classical $\alpha$ effect, which
is traditionally believed to operate in galaxies.
All four flows have the following $x$ and $y$ components:
\begin{tbequation}
u_x =  v_0 \, \sin k_0x \, \cos k_0y, \quad
u_y = -v_0 \, \cos k_0x \, \sin k_0y,
\label{KH21}
\end{tbequation}
but the $z$ components are different for each flow:
\begin{tbequation}
u_z=w_0 \left\{
\begin{array}{ll}
\sin k_0x \, \sin k_0y
\quad\mbox{(for flow I)},
\\
\cos k_0x \, \cos k_0y
\quad\mbox{(for flow II)},
\\
(\cos 2 k_0x + \cos 2 k_0y)/2
\quad\mbox{(for flow III)},
\\
\sin k_0x
\quad\mbox{(for flow IV)},
\label{KH29}
\end{array}
\right.
\end{tbequation}
where $v_0$, $w_0$, and $k_0$ are constants.
Particular solutions are obtained by specifying the
magnetic Reynolds number $\Rm=v_0/\eta k_0$ and
the ratio $w_0/v_0$.
The magnetic field must always be 3-D and
varies in the $z$ direction like $e^{\ii k_z z}$, where
$k_z$ is sometimes chosen such that it maximizes the growth rate.
\end{textbox}

We summarize the essential features of flows~I--IV in \Tab{TRob}.
The resulting mean fields for flow~I can be interpreted in terms of an
$\alpha$ effect; see the text box on ``The $\alpha$ effect''.
The mean field for flow~IV was identified to be due to a
negative turbulent magnetic diffusivity \citep{DBM13}.
The origin of the mean field for flows~II and III involves the combination
of two different effects: turbulent pumping, which acts like an advection
velocity without actual material motion, and a memory effect, which means
that the electromotive force also involves the mean magnetic field from
earlier times.

These classifications can be formalized once we define an averaged magnetic
field $\meanBB$, which can here be an $xy$ planar average, so
$\meanBB=\meanBB(z,t)$ depends just on time and on one spatial coordinate.
This defines what we call the fluctuating field $\bb\equiv\BB-\meanBB$.
For the Roberts flows, there is no mean flow, i.e., $\meanUU=0$,
so the evolution of $\meanBB$ is only governed by the mean electromotive force
$\meanEMF\equiv\overline{\uu\times\bb}$, consisting of fluctuations only.

In all cases, the mean magnetic field along the $z$ axis,
$\meanBB_\|\equiv (0,0,\meanB_z)$, vanishes.
The perpendicular components, $\meanBB_\perp\equiv(\meanB_x,\meanB_y,0)$
are finite and we only need to focus on the components $\meanEMF_\perp$,
$\meanBB_\perp$, and $\meanJJ_\perp$.
For flow~I, which is maximally helical, there is a systematic swirl. 
As we have explained in the text box on ``The $\alpha$ effect,'' as a
result of this systematic swirl, flow~I produces an $\alpha$ effect,
and thus, we have
\begin{equation}
\meanEMF=\alpha\meanBB-\etat\mu_0\meanJJ.
\label{meanEMF}
\end{equation}
In \Eq{meanEMF}, $\etat$ is the turbulent magnetic diffusivity,
because it adds to the microphysical magnetic diffusivity $\eta$ to give
the total magnetic diffusivity $\etaT=\etat+\eta$.
For flows~II and III, the situation is more complicated in that $\alpha$
is now a tensor with vanishing diagonal components.
For flow~IV $\alpha$ is zero and $\etat$ is negative, which can thus
lead to exponential growth.
For all those flows, it is important to realize that $\alpha$ and
$\etat$ are in general scale-dependent, and $\etat$ becomes positive
when $\meanBB(z,t)$ has high spatial Fourier components, i.e., for mean
fields of smaller scale in the $z$ direction.
The dependence of $\meanEMF$ on the mean magnetic field $\meanBB$ and
its associated mean current density, $\meanJJ=\nab\times\meanBB/\mu_0$,
is discussed below.

\begin{marginnote}[]
\entry{Scale dependence}{$\alpha$ and $\etat$ decrease toward smaller scales}
\end{marginnote}

To determine all components of the tensors
$\alpha_{ij}$ and $\eta_{ijk}$ in the representation
$\meanemf_i=\alpha_{ij}\meanB_j+\eta_{ijk}\meanB_{j,k}$ with a rank three
tensor $\eta_{ijk}$, one must solve the equation for the fluctuations
in terms of the mean magnetic field.
Here, a comma denotes partial differentiation.
In \Tab{TRob}, we indicate the form of $\meanEMF_\perp$ for each of the
four flows.
\begin{marginnote}[]
\entry{Magnetic Reynolds number}{Defined here as $\Rm=\urms/\eta\kf$,
where $\kf$ is the wavenumber of velocity fluctuations}
\end{marginnote}
We also indicate the critical values of the magnetic
Reynolds number $\Rmc$, above which dynamo action occurs.
Here, $\Rmc$ is defined with $\kf=k_0$, and we have fixed $w_0=v_0$
and $k_z=k_0/2$ to ensure that dynamos are possible for all four flows.
For flows~II and III, for example, no dynamos are possible for $k_z=k_0$.

\begin{table}[t]
\tabcolsep7.5pt
\caption{
Robert flows~I--IV as simple benchmarks, and their dynamo properties.
}
\label{TRob}
\begin{center}
\begin{tabular}{@{}c|c|c|c|c@{}}
\hline
Flow & helicity & interpretation & $\meanEMF_\perp$ & $\Rmc$ \\ 
\hline
I   & yes, and constant & $\alpha$ effect & $\alpha\meanBB_\perp-\etat\meanJJ_\perp$ & 1.99 \\
II  & pointwise zero $\alpha$ & \makecell{off-diagonal $\alpha$ tensor \\ with memory effect}
    & $\pmatrix{0&a\cr a&0}\meanBB_\perp-\etat\meanJJ_\perp$ & 6.86 \\
III & zero only on average    & \makecell{pumping effect \\ with memory effect}
    & $\pmatrix{0&\gamma\cr-\gamma&0}\meanBB_\perp-\etat\meanJJ_\perp$ & 3.92 \\
IV  & zero only on average & \makecell{negative ``turbulent'' \\ diffusion}
    & \makecell{$-\etat\meanJJ_\perp$ with $\etat<0$ \\ on large length scales} & 4.55 \\
\hline
\end{tabular}
\end{center}
\begin{tabnote}
In all cases, $k_z=k_0/2$ was used.
The values of $\urms$ are 0.866 for flows~I--III and unity for flow~IV.
\end{tabnote}
\end{table}

\subsection{Large-scale dynamos and averaging}
\label{Averaging}

As alluded to above, an important feature of the four Roberts
flow dynamos is that all of them are examples of large-scale dynamos,
i.e., one can define an average (here an $xy$ average) under which
the magnetic field retains most of its energy and still captures its
essential spatio-temporal evolution.
The most suitable type of averaging depends on the type of the mean
magnetic field that can emerge in certain geometries and in certain
parameter regimes; see \cite{Gent2013} and \cite{Hollins+22} for a
discussion.
For example, in the context of disk galaxies, the azimuthally averaged
magnetic field plays an important role.
In cylindrical coordinates, $(\varpi,\phi,z)$, such a field depends --
not necessarily smoothly -- on the cylindrical radius $\varpi$ and the
height $z$ above the midplane, as well as on time.
This dependence may still involve rapid variability, which can easily
lead to a confusing terminology when we want
to split the magnetic field into mean fields and fluctuations, $\BB=\meanBB+\bb$.
To avoid the temptation to refer to the non-smoothness of $\meanBB$ as
fluctuations, one sometimes refers to ordered and random fields instead
(SS22).

An azimuthal average has obviously no azimuthal dependence and cannot
describe nonaxisymmetric magnetic fields.
On the other hand, in a mean-field model, one can always just assume
that $\meanBB$ also depends on $\phi$.
This mean field could be understood as a low Fourier mode filtering.
However, then the average of the product of a mean and a fluctuation
vanishes only approximately; see \cite{ZBC18} for the related discussion
on what is known as Reynolds rules for averaging.

Regarding the periodic flow patterns in Cartesian coordinates discussed
in \Sec{SimpleDynamos}, it is important to stress that there can be
examples where planar $xy$ averaging is not suitable.
An example is the Taylor--Green flow, where a one-dimensional average
(here a $z$ average) must be taken to demonstrate the existence of a
large-scale dynamo due to a negative turbulent magnetic diffusivity
\citep{ABNZ15}.
In that case, the mean field depends on $x$, $y$, and $t$.

\subsection{Types of large-scale dynamos}

Historically, the $\alpha$ effect was the first distinct dynamo effect
that was discovered.
It emerged in the derivation of mean-field effects in stratified
rotating turbulence \citep{SKR66}, but in its essence, it was
already obtained by \cite{Par55} using a more phenomenological approach.
It is intrinsically connected with the presence of kinetic helicity
and is proportional to the pseudo-scalar $\grav\cdot\OO$,
as discussed in the text box on ``The $\alpha$ effect.''
Dynamos can work with an $\alpha$ effect alone, in which case one
talks about an $\alpha^2$ dynamo.
Astrophysical dynamos often have strong shear, so there is an extra
$\meanUU\times\meanBB$ term on the right-hand side of \Eq{dAmeandt},
but shear alone cannot produce a dynamo.
When shear is complemented by an $\alpha$ effect, one talks about an
$\alpha\Omega$ dynamo, or even an $\alpha^2\Omega$ dynamo if one wants
to emphasize that both $\alpha$ and $\Omega$ effects play a role.

We do not know whether galactic dynamos are of $\alpha\Omega$ type.
Alternatives include the incoherent $\alpha$--shear effect, but also the
(magnetic) shear--current effect has been discussed \Sec{ShearCurrent}; see \Tab{TDynType}
for a summary of the different types of large-scale dynamos known so far.
Here we also indicate whether a small-scale dynamo might operate and
whether the dynamo is expected to be fast, i.e., to grow even for very
large values of $\Rm$.
This is usually not the case for laminar flows, unless the flow
has chaotic streamlines.\footnote{The Galloway--Proctor flow
\citep{GP92} is an example of a laminar flow that is fast.
It is a Roberts flow with time-dependent phases in the trigonometric
functions, which causes its streamlines to be chaotic.}

\begin{table}
\tabcolsep7.5pt
\caption{
Summary of different types of large-scale dynamos.
}
\label{TDynType}
\begin{center}
\begin{tabular}{@{}l|c|c|c@{}}
\hline
Flow & main dynamo effect & small-scale & fast \\
\hline
helical turbulence                & $\alpha^2$ dynamo & \checkmark & \checkmark \\
Roberts flow I (laminar, helical) & $\alpha^2$ dynamo &    ---     &    ---     \\
Roberts flows II and III (laminar, nonhelical) & time delay &    ---     &    ---     \\
Roberts flows IV                  &neg turb diff&   ---     &    ---     \\
R\"adler effect with shear        &$\OO\times\meanJJ$ effect&    ---    & \checkmark \\
(magnetic) shear--current effect  &$\SSSS\meanJJ$ effect    &    ---    & \checkmark \\
incoherent $\alpha$--shear effect &fluctuating $\alpha$ effect& \checkmark & \checkmark \\
\hline
\end{tabular}
\end{center}
\begin{tabnote}
`Small-scale' refers to possibility that a small-scale dynamo would operate
together with a large-scale dynamo.
`Fast' refers to the possibility that the dynamo works in the limit
$\eta\to0$, as discussed in \Sec{NeedMagneticDiffusivity}.
\end{tabnote}
\end{table}

\subsubsection{Helical dynamos}

Roberts flow~I is maximally helical.
It is a prototype of an $\alpha^2$ dynamo, whereby the two nonvanishing
horizontally averaged mean-field components, $\meanB_x$ and $\meanB_y$,
are being amplified by the $\alpha$ effect.
If shear is important, and we have an $\alpha\Omega$ dynamo,
the dynamo is often oscillatory and can exhibit traveling wave
solutions.
In oblate bodies such as galaxies, however, $\alpha\Omega$ dynamos
are usually non-oscillatory \citep{Par79,Stix75}.
\begin{marginnote}[]
\entry{Roberts flow I}{is a prototype of an $\alpha^2$ dynamo;
the flow is fully helical}
\end{marginnote}

\subsubsection{Nonhelical large-scale dynamos}

There are various examples of large-scale dynamos that do not involve
magnetic helicity.
Three of the four Roberts flows have clearly demonstrated that large-scale dynamos
do not have to be helical and they can even have pointwise zero helicity.
Common to all three examples of Roberts flows II--IV is the fact that
the two components, $\meanB_x$ and $\meanB_y$, are uncoupled from
each other.
In these examples, the two components have the same growth rate, but there are other
flows, such as the Willis flow \citep{Wil12}, where the growth rates
of $\meanB_x$ and $\meanB_y$ are different and only one of the two
components grows.
This is unusual and different from conventional dynamos of $\alpha\Omega$
or $\alpha^2$ type, where the two components have strictly the same
growth rate.
Mathematically, the coupling of the two mean field components is caused
by the cross product in the expression $\nab\times(\alpha\meanBB)$
on the right-hand side of the evolution equation for $\meanBB$.
In the presence of shear, for example by a mean flow with constant
shear $S=\partial\meanU_y/\partial x$, one has
$\partial\meanB_y/\partial t=S\meanB_x+...$, where the ellipsis denotes
further terms not relevant to the present discussion.
\begin{marginnote}[]
\entry{Pumping velocity $\bm{\gamma}$}{is a contribution to the mean
electromotive force given by the off-diagonal terms of the $\alpha$
tensor through $\gamma_i=-\frac{1}{2}\epsilon_{ijk}\alpha_{jk}$}
\end{marginnote}

The reason for the decoupling of the two magnetic field components in 
some examples is that the dynamo-active
terms operate on each field component separately (i.e.,
$\partial\meanB_x/\partial t=-\ggamma\cdot\nab\meanB_x$ and
$\partial\meanB_y/\partial t=-\ggamma\cdot\nab\meanB_y$
for dynamos where the pumping velocity $\ggamma$ has a memory effect).
In its simplest form, a memory effect has an exponential kernel
proportional to $e^{-(t-t')/\tau}$ for $t>t'$, and zero otherwise.
Here, $t$ is the current time and $t'$ the integration variable,
covering all earlier times.
In Fourier space, it leads to a factor
$1/(1-\ii\omega\tau)$, where $\omega$ is the
frequency and $\tau$ is the turnover time.
When $\gamma\omega\tau>(\eta+\etat)k$, dynamo action becomes possible.

\subsubsection{R\"adler and shear--current effects}
\label{ShearCurrent}

Early in the history of mean-field dynamo theory, \cite{Radler69} found a novel
large-scale dynamo effect for rotating, but unstratified bodies, whereby
$\meanEMF$ has a term proportional to $\OO\times\meanJJ$.
Here, $\OO$ is a pseudovector pointing along the rotation axis.
The azimuthal velocity is then $\uu_\phi=\pom\times\OO$.
However, it is easy to see that the $\OO\times\meanJJ$ term in $\meanEMF$
does not contribute to the generation of mean-field energy proportional
to $\meanBB^2$, because the dot product with $\meanJJ$ vanishes.
Therefore, additional effects are needed to achieve dynamo action.
Shear is one such effect, which can also generate another large-scale
dynamo, similar to the  R\"adler effect: the shear--current effect.
Most of the numerical evidence today shows that this effect does not
have a favorable sign for dynamo action \citep{BRRK08}.
There is the possibility that this finding would change when 
the shear-current effect
is strongly controlled by the
magnetic field from a small-scale dynamo \citep{SB15c}.
While it is true that large-scale magnetic fields can be generated, it
is possible that the real reason behind this is actually the incoherent
$\alpha$ effect, as will be discussed below; see also \cite{ZB21}
for a detailed assessment of the different possibilities.

\subsubsection{Incoherent and shear dynamo effects}

Another important class of large-scale dynamos may explain
the phenomenon of large-scale magnetic field generation in shear flows
without helicity.
Such nonhelical dynamo action was first found in a more complex shear
flow geometry, relevant to the solar tachocline at mid to low latitudes
\citep{Bra05}.
In this environment, large-scale fields can be generated
both with and without helicity in the driving of the turbulence.
Subsequent studies by \cite{You+08} and \cite{BRRK08} produced such
dynamos in simpler shearing box simulations, but gave different
interpretations, which we discuss below.

One interpretation involves helicity fluctuations, which lead to
an incoherent $\alpha$ effect and, in conjunction with shear,
to large-scale dynamo action \citep{VB97}.
An incoherent $\alpha$ effect can lead to a negative turbulent magnetic
diffusivity \citep{Kra76}.
In that sense, the incoherent $\alpha$ effect is actually similar to
the dynamo effect in Roberts flow IV.
\begin{marginnote}[]
\entry{Incoherent $\alpha$ effect}{an $\alpha$ effect with frequent sign changes}
\end{marginnote}

Another interpretation is what is sometimes called the shear dynamo.
Attempts to interpret this as a mean-field effect amounts to invoking
the shear-current effect.
The magnetic shear current effect, by contrast, is based on correlated
fluctuations of the magnetic field from a small-scale dynamo, which is
assumed to operate in the background.

The role of the incoherent $\alpha$ effect in galactic dynamos is
uncertain and may have been underestimated in the past.
It might be important if the net kinetic helicity above and below
the midplane is small.
This may well be the case, especially when there is significant
interaction with the CGM.
Such interactions can generate strong fluctuations of opposite sign in
the kinetic helicity, which would cancel out.

\subsection{Small-scale dynamos}

Under fully isotropic conditions and without helicity, dynamo action is
still possible---both for large and small values of the magnetic Prandtl
number $\Pm$ \citep{Kaz68}; see the detailed discussion by \cite{Scheko+04}.
The existence of small-scale dynamos under isotropic conditions implies
that the concept of nonmagnetic Kolmogorov turbulence hardly exists in
astrophysics, where the medium is usually always highly conducting.

\subsubsection{Early work on the subject}

\begin{figure}[t]
\includegraphics[width=\textwidth]{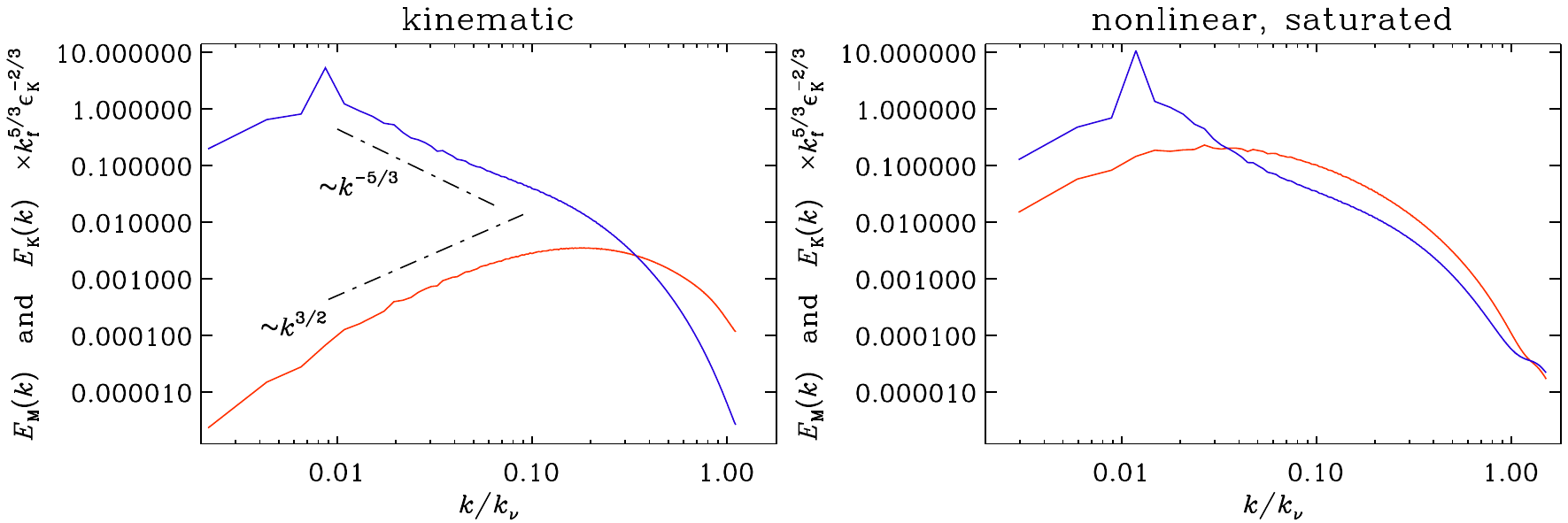}
\caption{
Magnetic (red lines) and kinetic (blue lines) energy spectra
during the kinematic (left) and nonlinear saturated (right) phases.
Here, $k_\nu=(\epsK/\nu^3)^{1/4}$ with $\epsK$ being the dissipation rate.
Note how the peak of $\EM(k)$ shifts to larger scales in
the saturated case.
Figure adapted from Run~E of \cite{Bra+22}.
}\label{pspec_comp_kinsat_Pm1_kf4_rev}
\end{figure}

In the early {\em kinematic} regime, when the magnetic field is still weak and
exponentially growing, its energy spectrum increases with wavenumber
$k$ proportional to $k^{3/2}$ and has a peak at the resistive
wavenumber, provided $\Pm\gg1$.
\cite{KA92} found that the peak occurs at a wavenumber $k_\eta$ that depends on
the growth rate $\lambda$ through $k_\eta=\sqrt{4\lambda/15\eta}$.
In the {\em saturated} stage, the peak of the magnetic spectrum shifts
closer to the forcing scale; see \Fig{pspec_comp_kinsat_Pm1_kf4_rev}.
More recent work on small-scale dynamos is numerical, and is covered
in detail in the following sections.

\subsubsection{Effect of ambipolar diffusion}

\cite{XL16} found a strong similarity between the regime of large magnetic
Prandtl numbers and the regime of partial ionization.
Their results have been confirmed in numerical simulations \citep{Xu+19}.
In those simulations, the microphysical magnetic Prandtl number remained
undetermined, because no explicit viscosity or magnetic diffusivity
were used.
Two-fluid direct numerical simulations \citep{Bra19} showed that at
large magnetic Prandtl number, the kinetic energy spectra for neutrals
and ions show different slopes.
The energy spectra of ions and neutrals depart from each other only a
small scales when $k/k_\nu>1$.
For larger ambipolar diffusion coupling, the kinetic energy spectra of
neutrals decrease further while those of the ions increase slightly.

\section{CATASTROPHIC QUENCHING AND MAGNETIC HELICITY FLUXES}

As long as the magnetic field is weak, the Lorenz force
plays no significant role.
Many dynamo effects, including those discussed in \Sec{SimpleDynamos},
can then be fully described by a given velocity field.
However, as soon as the velocity field is determined or modified by the
magnetic field, the dynamo problem becomes nonlinear.
Eventually, the growing effect of the Lorenz force on the flow can limit
(or quench) the magnetic field growth.

\subsection{Catastrophic quenching for uniform magnetic fields}
\label{Catastrophic_uniform}

The term catastrophic quenching was coined by \cite{BF00} to denote any
type of detrimental $\Rm$ dependence of the nonlinear feedback.
Ignoring the effect of turbulent diffusion, i.e., the term
$-\etat\mu_0\meanJJ$ in \Eq{meanEMF}, \cite{CH96} found numerically
that $\alpha\propto(1+\Rm\meanBB^2/\Beq^2)^{-1}$, where
$\Beq=\sqrt{\mu_0\rho_0}\urms$ is the equipartition field strength,
whose energy density is equal to the kinetic energy density.
Evidently, owing to the $\Rm$ factor in the expression for $\alpha$,
this dependence is ``catastrophic''.
This dependence was originally anticipated
by \cite{VC92} based on earlier analogous results by \cite{CV91} for
the suppression of just $\etat$ in 2-D.
\cite{GD96} explained these results as a consequence of the
conservation of magnetic helicity $\bra{\AAA\cdot\BB}$ in 3-D,
which is routinely seen during laboratory plasma relaxation \citep{Ji+95}.
However, the dependence of $\alpha$ on $\Rm$ is
a peculiar property of the magnetic helicity
equation in the presence of an imposed magnetic field $\BB_0$.
In that case, the magnetic helicity corresponding to the departure
from the imposed field, $\bb$, yields, in the steady state,
$0=\bra{(\uu\times\BB_0)\cdot\bb}-
\eta\mu_0\bra{\jj\cdot\bb}$.
Since we define here mean fields as volume averages, and since
$\bra{\JJ}=0$ in \Eq{meanEMF}, we have
$\bra{(\uu\times\BB_0)\cdot\bb}
=-(\bra{\uu\times\bb})\cdot\BB_0=-\alpha\BB_0^2$,
and therefore
$\alpha=-\eta\mu_0\bra{\jj\cdot\bb}/\BB_0^2$,
so $\alpha\to0$ as $\eta\to0$ or $\Rm\to\infty$.
This agrees with the heuristic quenching formula
$\alpha\propto(1+\Rm\meanBB^2/\Beq^2)^{-1}$, which also
predicts $\alpha\to0$ as $\Rm\to\infty$.
The analysis also shows that the quenching is related to
magnetic helicity conservation.
A detailed explanation of this derivation is reviewed in \cite{BS05}.

Much of the original work on catastrophic quenching adopted
periodic domains.
This is clearly only of limited value when thinking about galaxies.
This result for $\alpha$ in the nonlinear regime was first obtained
by \cite{Keinigs+83}.
However, it is not really relevant in practice, because it assumes that
the magnetic field can meaningfully be described by volume averages.
This is not the case, because a volume-averaged magnetic field is always
constant in a periodic domain.

A relevant mean field for this kind of problem can be defined as 
planar averages, as discussed in \Sec{SimpleDynamos}.
We denote that by overbars.
The diffusion term $\etat\mu_0\meanJJ$ can then not be neglected and
the relation of \cite{Keinigs+83} can then be written in the form
$\alpha-\etat k_{\rm m}=-\eta\mu_0\bra{\jj\cdot\bb}/\BB_0^2$, where
$k_{\rm m}=\mu_0\meanJJ\cdot\meanBB/\meanBB^2$.
This would mean that only the difference $\alpha-\etat k_{\rm m}$,
not $\alpha$ itself, is quenched catastrophically.

\subsection{Catastrophically slow saturation in closed domains}
\label{Catastrophic_closed}

In reality, even if the restriction to closed or periodic domains is
retained, neither $\alpha$ nor $\etat$ are quenched in a catastrophic
fashion \citep{BRRS08}.
Instead, the timescale for reaching ultimate saturation is
``catastrophically'' prolonged, i.e., the final saturation obeys
\citep{Bra01}
\begin{equation}
\meanBB^2\approx\bra{\bb^2}\,(\kf/k_1)\,
\left[1-e^{-2\eta k_1^2(t-t_{\rm sat})}\right]
\quad\mbox{(for $t>t_{\rm sat}$)},
\label{FinalSat}
\end{equation}
where $\kf$ is the typical wavenumber of the turbulence, $k_1=2\pi/L$
is the lowest wavenumber of the cubic domain of size $L^3$, and
$t_{\rm sat}$ marks the end of the early kinematic growth phase and
the beginning of the slow saturation phase.
Let us emphasize once again that in \Eq{FinalSat}, the value of $\eta$
is the microphysical value, which is extremely small in galaxies.
This motivates the characterization as ``catastrophically slow''.

\begin{textbox}[t]\section{Derivation of \Eq{FinalSat}}
In periodic domains,
the slow saturation behavior after $t=t_{\rm sat}$ is governed by magnetic
helicity conservation.
The uncurled induction equation reads,
$\partial\AAA/\partial t=-\EE-\nab\varphi$, where
$\EE=-\UU\times\BB+\eta\mu_0\JJ$ is the electric field
and $\varphi$ is the electrostatic potential.
The evolution of the magnetic helicity density $\AAA\cdot\BB$ is then given by
\begin{tbequation}
\frac{\partial}{\partial t}(\AAA\cdot\BB)=\cancel{2(\UU\times\BB)\cdot\BB}
-2\eta\mu_0\JJ\cdot\BB-\nab\cdot\FFFF,
\label{dABdt}
\end{tbequation}
where $\FFFF=\EE\times\AAA+\varphi\BB$ is the magnetic helicity flux density.
(Note the analogy with the Poynting flux $\EE\times\BB/\mu_0$ of magnetic
energy density.)
The equations involving $\AAA$ and $\FFFF$ depend on the gauge, i.e.,
on the form of $\varphi$, which can be chosen freely.
One frequently adopts the Weyl gauge, $\varphi=0$.

Next, we consider spatial averages $\meanBB=\nab\times\meanAA$ and
$\meanJJ=\nab\times\meanBB/\mu_0$, along with the resulting fluctuations,
$\aaaa=\AAA-\meanAA$, $\bb=\BB-\meanBB$, and $\jj=\JJ-\meanJJ$, so,
after averaging, \Eq{dABdt} becomes
\begin{tbequation}
\frac{\partial}{\partial t}(\meanAA\cdot\meanBB+\overline{\aaaa\cdot\bb})
=-2\eta\mu_0\left(\meanJJ\cdot\meanBB+\overline{\jj\cdot\bb}\right)
-\nab\cdot\meanFFFFm,
\label{dAmBmdt}
\end{tbequation}
where $\meanFFFFm$ is the magnetic helicity flux for the mean field.
Our analysis concerns only the phase when the small-scale
dynamo has already saturated (for $t>t_{\rm sat}$), so
$\overline{\aaaa\cdot\bb}$ is approximately constant in time.
Assuming the field to be helical with negative helicity at small
scales and positive at large scales, we have
$\mu_0\overline{\jj\cdot\bb}\approx-\kf\overline{\bb^2}$ and
$\meanAA\cdot\meanBB\approx\overline{\BB^2}/k_1
\approx\mu_0\meanJJ\cdot\meanBB/k_1^2$.
Inserting this into \Eq{dAmBmdt} and performing volume averaging
over the whole domain, indicated by angle brackets, so
that the flux divergence term vanishes, one obtains
\begin{tbequation}
\frac{\dd}{\dd t}\bra{\meanBB^2}
=-2\eta k_1^2\bra{\meanBB^2}+2\eta k_1\kf\bra{\bb^2},
\label{dBm2dt}
\end{tbequation}
the solution of which for $\bra{\bb^2}=\const$ is given by \Eq{FinalSat}.
\end{textbox}

The derivation of \Eq{FinalSat} is based on just the magnetic helicity
equation, i.e., no mean field theory was invoked; see the
text box on the ``Derivation of \Eq{FinalSat}''.
However, a phenomenological mean field theory can be formulated where
the $\alpha$ effect has an extra magnetic contribution related to the
magnetic helicity at small scales, which, in turn, is computed based on
the large-scale magnetic helicity that is being being produced by the
mean-field dynamo under the assumption that the total magnetic helicity
is conserved.
The $\alpha$ effect itself is then not catastrophically quenched
\citep{BB02}, so the magnetic field (in a periodic or closed domain)
can still be strong, but only after a resistively long time;
see, again, \cite{BS05} for a review.

\subsection{Alleviating catastrophic quenching by magnetic helicity fluxes}
\label{sec:helicityflux}
It has long been hypothesized that the action of magnetic helicity
fluxes can overcome what would otherwise be an extremely slow approach to
saturation (as described in \Sec{Catastrophic_closed}) or a saturation
at a very low amplitude \citep{GD96}.
The latter has been demonstrated in \cite{BDS02}, who discussed 
the preferential removal of small scale magnetic fields.
In this experiment, they periodically removed magnetic field at high Fourier
modes from the simulation.
After each removal, the small-scale field was then no longer saturated and
was thus allowed to grow and regain old strength, while the large-scale
field grew to larger strength than before.
This continued with each removal step.
While the idea is simple and convincing, there is as yet no conclusive
demonstration from simulations that this also works with actual magnetic
helicity fluxes.

For assessing the role of magnetic helicity fluxes,
the decisive equation is that for the magnetic helicity of the
fluctuating field, $\overline{\aaaa\cdot\bb}$.
The fluctuating field can be determined from the equation for the
mean field, which, under the Weyl gauge, can be written as
\begin{equation}
\frac{\partial\meanAA}{\partial t}=\meanUU\times\meanBB+\meanEMF
-\eta\mu_0\meanJJ,
\label{dAmdt}
\end{equation}
where we recall that $\meanEMF\equiv\overline{\uu\times\bb}$
is the mean electromotive force.
This expression results in the following equation for the magnetic
helicity of the mean magnetic field:
\begin{equation}
\frac{\partial}{\partial t}(\meanAA\cdot\meanBB)=2\meanEMF\cdot\meanBB
-2\eta\mu_0\meanJJ\cdot\meanBB-\nab\cdot\meanFFFFm.
\label{dAmBmdt2}
\end{equation}
The equation for $\overline{\aaaa\cdot\bb}$ must also have a corresponding
$\meanEMF$ term, $-2\meanEMF\cdot\meanBB$,
\begin{equation}
\frac{\partial}{\partial t}
\overline{\aaaa\cdot\bb}=-2\meanEMF\cdot\meanBB
-2\eta\mu_0\overline{\jj\cdot\bb}-\nab\cdot\meanFFFFf,
\label{dabdt}
\end{equation}
so that the sum of both equations yields \Eq{dABdt}.
Here, $\meanFFFFf$ is the magnetic helicity flux for the fluctuating field.

\begin{figure}[t]
\includegraphics[width=.8\textwidth]{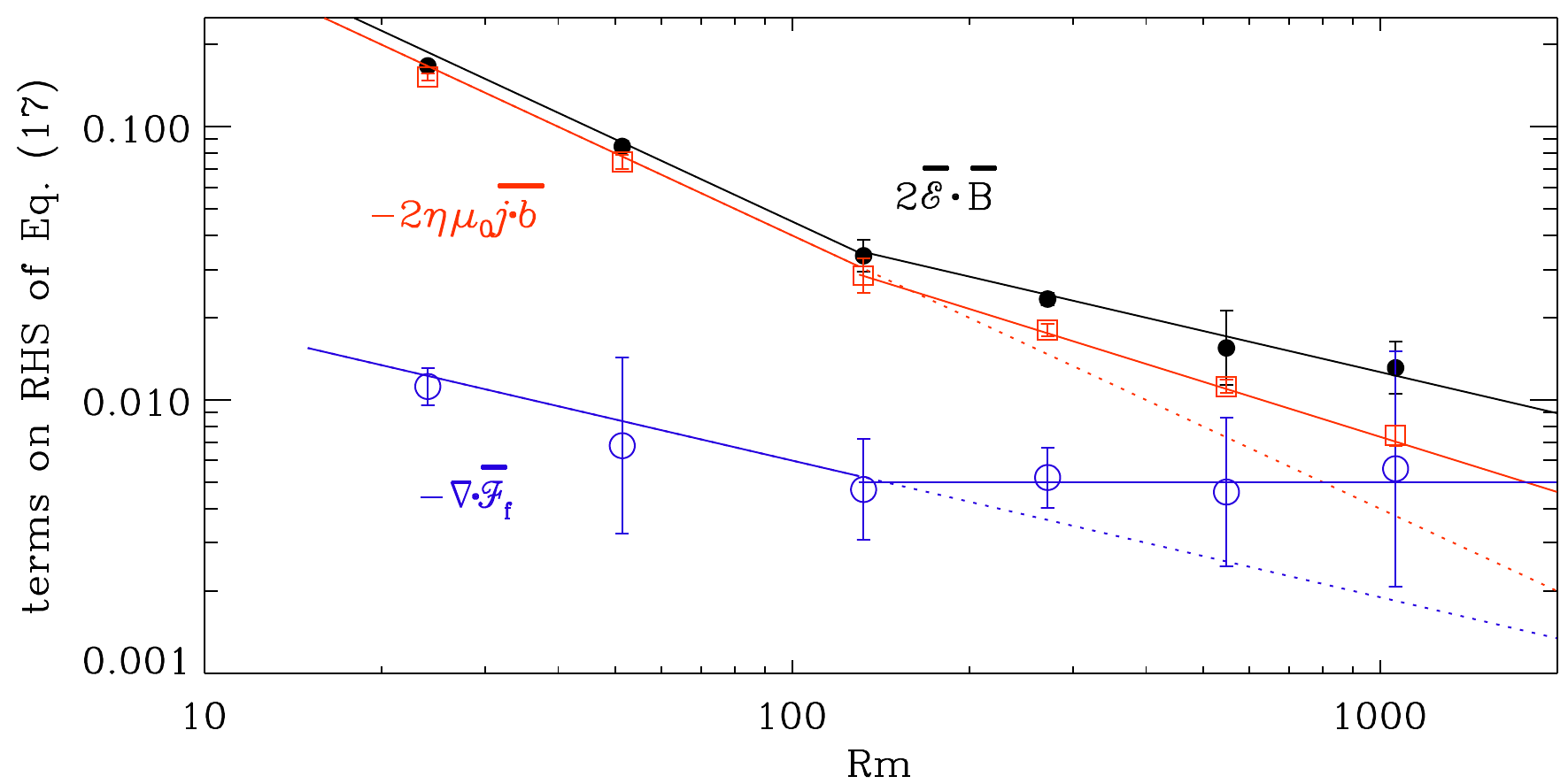}
\caption{
$\Rm$ dependence of terms on the right-hand side of the small-scale
magnetic helicity equation \eq{dabdt}.
Note that $\nab\cdot\meanFFFFf$ becomes comparable to $2\meanEMF\cdot\meanBB$
and $2\eta\overline{\jj\cdot\bb}$ only for $\Rm>1000$.
Adapted from \cite{DSGB13}.
}\label{presults2b}
\end{figure}

In the steady state, there are three terms on the right hand side,
$2\meanEMF\cdot\meanBB$, $2\eta\mu_0\overline{\jj\cdot\bb}$, and
$\nab\cdot\meanFFFFf$.
Simulations by \cite{DSGB13} and \cite{Rin21} showed that 
the helicity flux divergence begins to become more important than the resistive terms
only at $\Rm$ of the order of 1000 (see \Fig{presults2b}).
Both works showed the presence of turbulent diffusive magnetic helicity
fluxes in the simulations.
Those fluxes were proportional to the negative gradient
of the local magnetic helicity density.
In the work of \cite{DSGB13}, there was also a galactic wind contributing
to an advective magnetic helicity flux proportional to the wind speed.
One could expect the saturation behavior to become independent of $\Rm$.
However, simulations still show that $\meanBB^2$ declines with increasing $\Rm$.
This could mean that $\Rm$ needs to be much larger than 1000, but
probing this regime requires larger simulations.
It remains then to be seen if future simulations with different setups
can result in situations where $2\eta\mu_0\overline{\jj\cdot\bb}$ does
become clearly subdominant.

\section{MEAN-FIELD COEFFICIENTS AND NONLOCALITY}

\subsection{Parameterization of the mean electromotive force}
\label{ParameterizationEMF}

The mean electromotive force, $\meanEMF$, in \Eq{dAmdt} can be
expressed nonlocally in terms of the mean magnetic field as
\begin{equation}
\meanemf_i=\alpha_{ij}\ast\meanB_j+\eta_{ijk}\ast\partial\meanB_j/\partial x_k,
\label{EMFi}
\end{equation}
where the asterisks denote a convolution over space and time, and
$\alpha_{ij}$ and $\eta_{ijk}$ are integral kernels and $x_k$ is the
$k$th component of the spatial coordinate, i.e.,
$\partial\meanB_j/\partial x_k=\meanB_{j,k}$.
For planar averages that depend on just one direction, we can write
$\meanemf_i=\alpha_{ij}\ast\meanB_j-\eta_{ij}\ast\meanJ_j$,
where $\alpha_{ij}$ and $\eta_{ij}$ would each only have four components.
For the rest of this review, we restrict ourselves to this simpler case,
but we refer the reader to \cite{Warnecke+18} for a study in the context
of 3-D convection in a sphere.

Most of the published literature ignores the fact that $\alpha_{ij}$
and $\eta_{ij}$ are integral kernels, and approximates the convolution
by a multiplication.
This approximation then assumes a local connection between $\meanEMF$
and the mean fields.
It ignores the effect of strong variations of the mean field in space
and time.
In Fourier space, the convolution in \Eq{EMFi} becomes a multiplication,
so it describes the combined response of all Fourier modes.
This becomes relevant when measuring the mean-field coefficients
for sinusoidal mean fields; see \Sec{UsingTestFields}.
\begin{marginnote}[]
\entry{Convolution}{an operation that becomes a multiplication in Fourier space}
\end{marginnote}

\subsection{Mean-field coefficients}

One of the major advances in mean-field dynamo theory is the development
of numerical methods to avoid the limitations imposed by using analytic
approaches.
This concerns mainly the linearization of the evolution equations for
the magnetic and velocity fluctuations in a turbulent flow.

To obtain expressions for $\alpha_{ij}$ and $\eta_{ijk}$, one has to
solve the equations for the fluctuations $\uu$ and $\bb$.
The most important one is that for $\bb$ and is obtained by subtracting
the equation for $\meanBB$ from that for $\BB$.
The equations are nonlinear in the fluctuations.
In analytic approaches, those nonlinear terms are often ignored (SS22),
which is termed the second order correlation approximation,
but this restriction is no longer required in the numerical evaluations
of $\meanEMF\equiv\overline{\uu\times\bb}$.
This approximation is only valid when $\Rm\ll1$, or when the correlation time
is short (which is even for supernova-driven turbulence hardly the case).
Neither of the two is relevant to astrophysics, so we focus here
on a numerical, nonlinear approach, where no approximation is used.

When the linearization is abandoned, most of the changes in the
coefficients $\alpha_{ij}$ and $\eta_{ijk}$ are of quantitative nature,
especially when the mean field is weak.
There are a few examples where qualitatively new effects emerge: turbulent
pumping in the Galloway--Proctor flow, or the effect of kinetic helicity
on the turbulent magnetic diffusivity, although those effects remain
mainly of academic interest \citep[for a review, see][]{Bran18}.

\subsection{Methods for measuring $\alpha$ and other effects}

One approach is to use a nonlinear simulation to obtain $\uu$ and
$\bb$ in the presence of an additional imposed magnetic field.
The resulting $\overline{\uu\times\bb}$ can be related to $\meanBB$
by ignoring $\eta_{ij}$ and $\meanJJ$.
This is termed the imposed-field method, but it can only
be used when $\meanJJ$ vanishes, for example when the averages are
zero-dimensional, i.e., volume averages.

Another approach is to relate $\overline{\uu\times\bb}$ to the
actual $\meanBB$ and $\meanJJ$ by correlating them to each other and
computing $\alpha_{ij}$ and $\eta_{ij}$ as correlation coefficients.
This approach has been applied both for the integral kernels in the
nonlocal approach \citep{BS02,Bendre+22} and the coefficients in the
local version \citep{Simard+16}.
The reliability of this approach is unclear and it has not yet been
verified for the simple examples of the Roberts flow mean-field dynamos
discussed in \Sec{SimpleDynamos}.
This method is sometimes called the correlation method.
The occurrence of unphysical results with this method (e.g., $\etat<0$)
can sometimes be alleviated by using the singular value decomposition
\citep{Simard+16}.

The most reliable method for calculating $\alpha_{ij}$ and $\eta_{ij}$
is the test-field method (TFM), where one solves the equations for the
fluctuations numerically for a sufficiently big set of test fields.
In the following, we only describe its essence in a few words.
A more detailed description can be found in the review of \cite{Bran+10}.

\subsection{Using test fields}
\label{UsingTestFields}

The TFM was originally applied by \cite{Sch05,Sch07} to determine the
dependence of all transport coefficients in a sphere using longitudinal
averages.
In that case, one has 9 coefficients for $\alpha_{ij}$ and 18 nonvanishing
coefficients for rank three tensor $\eta_{ijk}$ in the representation
$\meanemf_i=\alpha_{ij}\meanB_j+\eta_{ijk}\meanB_{j,k}$.
(The nine coefficients $\eta_{ij\phi}$ do not enter the problem, because
$\phi$ derivatives of $\phi$ averages vanish.)
For systems in Cartesian coordinates, planar $xy$ averages are often
the most suitable; see \cite{Bra05} and \cite{BRRK08} for the
first applications.
The number of relevant coefficients is then four for
$\alpha_{ij}$, because only $i,j=1,2$ are relevant, and four
for the rank two tensor $\eta_{ij}$ in the representation
$\meanemf_i=\alpha_{ij}\meanB_j-\eta_{ij}\mu_0\meanJ_j$, because
there are only two nonvanishing components of $\meanB_{j,k}$ that can
be expressed as the two components of the mean current density with
$\mu_0\meanJ_x=-\meanB_{y,z}$ and $\mu_0\meanJ_y=\meanB_{x,z}$.
In that case, one can use four sinusoidal test fields $(\sin kz,0,0)$,
$(\cos kz,0,0)$, as well as $(0,\sin kz,0)$, $(0,\cos kz,0)$.

\begin{figure}[t]
\includegraphics[width=0.7\textwidth]{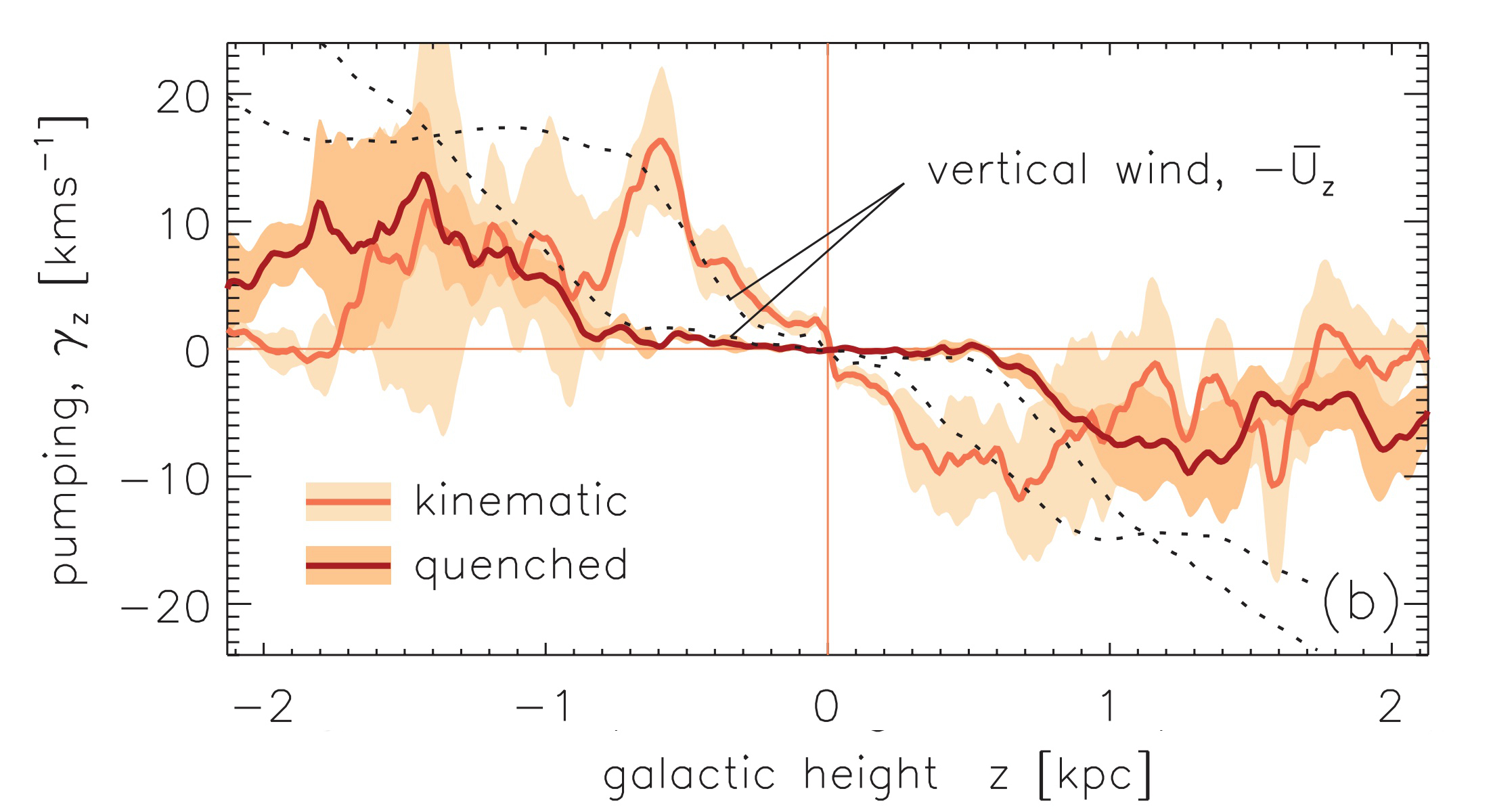}
\caption{
Dynamo coefficients from supernova-driven turbulence.
Shown here are the off-diagonal components, $\alpha_{\phi R}$ (black line)
and $-\alpha_{R\phi}$ (gray line), contributing to the pumping velocity
$\gamma_z=(\alpha_{\phi R}-\alpha_{R\phi})/2$.
The mean flow velocity is also shown.
Courtesy of \cite{Gressel+13}
}\label{Gressel}
\end{figure}

In \Fig{Gressel} we reproduce results from the work of
\cite{Gressel+08AN}, who performed simulations of dynamos from
supernova-driven turbulence in a portion of a stratified galactic disk.
Using the TFM with the Nirvana code, they found that $\etat$ increases away from the midplane
and that this leads to turbulent pumping toward the midplane, which is
given by $\ggamma\approx-(\tau/2)\nab\etat$.
The pumping velocity $\ggamma$ corresponds to off-diagonal components
of the $\alpha$ tensor, which they confirmed.
In particular, the pumping velocity in the $z$ direction is given by
$\gamma_z=(\alpha_{\phi R}-\alpha_{R\phi})/2$, where the subscript $R$
denotes cylindrical radius.
The pumping, also termed turbulent diamagnetism (SS22), pushes
the magnetic field toward the midplane and thereby strengthens the
dynamo \citep{BDMSST93}.
Surprisingly, this pumping increases toward smaller
scales \citep{Gressel+Elstner20}.

\subsection{Nonlocality in space and time}

It was soon realized that the results for $\alpha_{ij}$ and
$\eta_{ij}$ always depend on the wavenumber $k$.
This is explained in the box
``Evolution equation for nonlocality in space and time''.
For many turbulent flows, the components of both
$\alpha_{ij}$ and $\eta_{ij}$ decline with increasing values of $k$ in
a Laplacian fashion approximately proportional to $[1+(ak/\kf)^2]^{-1}$, where $a$
depends on details of the flow.
In this relation, the value of the empirical coefficient $a$
varied between 0.1 and 0.5, depending on the nature of the turbulent flow
\citep{RB12}.

The significance of nonlocality is that the transport coefficients
become effectively quenched when the mean field is of
small scale, i.e., smaller than the integral scale of the turbulence.
Especially near boundaries, where sharp boundary layers may occur in
calculations that ignore nonlocality, the actual field would be smoother.
In fact, sharp contrasting structures have been found in earlier galactic
dynamo simulations \citep{Moss96}.
Such results would need to be revisited in view of the importance of
nonlocality effects.

\begin{textbox}[t]\section{Evolution equation for nonlocality in space and time}
In Fourier space, the simplest empirical approximations to the spatial
and temporal nonlocalities, as obtained with the test-field method,
can be combined to a single expression, which reads
\begin{tbequation}
\tilde{\cal E}_i(k,\omega)=\frac
{\tilde{\alpha}_{ij}(k,\omega)\tilde{B}_j(k,\omega)
-\tilde{\eta}_{ij}(k,\omega)\mu_0\tilde{J}_j(k,\omega)}
{1-\ii\omega\tau+\ell^2k^2},
\label{EMFifrac}
\end{tbequation}
where $\ell=O(1/\kf)$ and $\tau=O(1/\urms\kf)$.
Moving the denominator to the left-hand side, the equation becomes
\begin{tbequation}
\left(1-\ii\omega\tau+\ell^2k^2\right)\tilde{\cal E}_i(k,\omega)=
\tilde{\alpha}_{ij}(k,\omega)\tilde{B}_j(k,\omega)
-\tilde{\eta}_{ij}(k,\omega)\mu_0\tilde{J}_j(k,\omega),
\label{EMFifrac2}
\end{tbequation}
which, back in real space, becomes a simple evolution equation with a
diffusion term on the right-hand side:
\begin{tbequation}
\tau\frac{\partial\meanemf_i}{\partial t}
=\alpha_{ij}\meanB_j-\eta_{ij}\mu_0\meanJ_j
+\ell^2\nabla^2\meanemf_i-\meanemf_i.
\label{EMFi_nonloc}
\end{tbequation}
This equation for the electromotive force is still only an approximation,
because there are in general also larger powers of $\omega$ and $k$,
but it provides a substantial improvement over the local formulations.
\end{textbox}

Even more important than spatial nonlocality is temporal nonlocality.
It is also termed a memory effect, because it implies that the
electromotive force depends not just on the magnetic field at the current
time, but also on the field at earlier times.
To leading order, the Fourier-transformed kernel of temporal nonlocality
is proportional to $(1-\ii\omega\tau)^{-1}$, where $\tau$ is the turbulent
turnover time.
Thus, the electromotive force diminishes with increasing frequency
$\omega$, but there is also a new imaginary component that was absent
otherwise.
This can lead to new dynamo effects such as that responsible in the
dynamos for the Roberts flows~II and III;
see the box box on ``Dynamos from the memory effect''.
Whether those effects play a role in turbulent dynamos is unclear.

Although the memory effect may not be strong enough to produce new dynamo
effects in turbulent flows, it is strong enough to produce significant
phase shifts between the generation of magnetic fields in galactic arm
and interarm regions.
This has been studied in detail by \cite{Shukurov98} and \cite{Chamandy+13}.
Including a memory effect in numerical simulations is, in general, very
cumbersome, because it requires storing the full spatial form of the
mean field for many earlier times in order to evaluate the convolution
integral in \Eq{EMFi}.
In the present case, however, and to leading order, the convolution integral
can be converted into an evolution equation for the electromotive
force, which is computationally much easier to solve; see \Eq{EMFi_nonloc}.
This approach was first proposed by \cite{RB12} and was applied to
dynamos in spheres \citep{BC18}.
This formalism also reproduces the dynamo effect from a time delay for
Roberts flows~II and III; see \Sec{SimpleDynamos}, as was demonstrated
by \cite{RDRB14}.
This is explained in the box ``Dynamos from the memory effect''.

\begin{textbox}[t]\section{Dynamos from the memory effect}

We emphasize, again, that dynamos from the memory effect are so far only
known to occur for the Roberts flow, so the effect may be special.
At this point, however, we can not exclude that the memory plays a role
in galaxies, for example in connection with the strong vertical
stratification leading to a pumping effect toward the midplane.
With the tools now at hand, it is now easy to explain this effect.

The dispersion relation for a problem with turbulent pumping
$\gamma$ and turbulent magnetic diffusion $\etat$ is given by
$\lambda=-\ii k\gamma-\etat k^2$.
Since $\Rey\lambda<0$, the solution can only decay, but it is
oscillating with the frequency $\omega=-\Imag\lambda=k\gamma$.
In the presence of a memory effect, $\gamma$ is replaced by
$\gamma/(1-\ii\omega\tau)$, where $\tau$ is the memory time.
Then, $\lambda\approx-\ii k\gamma\,(1-\ii\omega\tau)-\etat k^2$,
and $\Rey\lambda$ can be positive.
This is the case for the Roberts flow discussed in
\Sec{SimpleDynamos}.
\end{textbox}

\section{SETTING THE SCENE FOR DYNAMO ACTION IN REAL GALAXIES}

\subsection{Possibilities for seed magnetic fields}

The conditions in the early Universe provide several possibilities for 
seeding galactic dynamos. The seeds could be primordial, which generally 
means that they were generated during inflation or phase transitions,
or they could originate from a cosmic battery. Other theories also
involve later seeding from astrophysical processes. We examine these
possibilities in the following sections.
\begin{marginnote}[]
\entry{Cosmic battery}{A mechanism that can generate magnetic field even
when there is none initially}
\end{marginnote}

\subsubsection{The need for sufficiently strong seed magnetic fields}

In \Sec{sec:history} we calculated an average of 70 revolutions in the
Galaxy's lifetime.
This is not very much, so we have to be concerned about possible effects
on the strength and shape of the initial magnetic field.
Typical estimates for the growth rate of the Galactic dynamo are of the
order of $\Gamma\approx2\Gyr^{-1}$ \citep{Beck+96}.
This means that the mean magnetic field could be amplified by up to
12 orders of magnitude in about $14\Gyr$.
To reach the current level of the mean magnetic field of about $3\uG$,
we would need a seed magnetic field of about $10^{-18}\G$.
This is just about the level that can be expected from
the Biermann battery mechanism \citep{Rees87}.
The text box ``Battery mechanisms'' provides some more information about
the Biermann battery and other mechanisms that can generate magnetic
fields in an unmagnetized plasma.

Even though the growth rate of a galactic large-scale dynamo may just
be large enough for explaining the current level of the mean field
of $\approx3\uG$ at the present time, it would be insufficient for
explaining large-scale magnetic fields in very young (redshift $z=1$)
galaxies.
Observationally, however, such fields are believed to exist.
\cite{Kronberg+92} found evidence for strong magnetic fields in a
$z=0.395$ galaxy.
In a more systematic search, \citet{Bernet_2008} found strong RMs in
quasar sightlines passing from $z\simeq 1$ galaxy halos.
More recently, \citet{Mao2017} estimated a $\mu$G, kpc-coherent magnetic
field in a lensing galaxy at $z\simeq0.46$.
The fact that the RMs of the lensed images are similar provides
evidence for large-scale magnetic coherence.

To explain smaller-scale magnetic fields at equipartition levels of $3\uG$
would still require dynamo action, but that may just be a small-scale
dynamo.
Since the typical dynamo growth rates scale with the turbulence turnover
time, which is shorter at small scales, a small-scale dynamo is a viable
seed-field mechanism, as we discuss next.

\begin{textbox}[t]\section{Battery mechanisms}
\paragraph*{The Biermann battery.}
When the density and the temperature gradient in a plasma are misaligned,
the electrons move down the pressure gradient, generating an electromotive
force that gives rise to a magnetic field.
The resulting time derivative of the magnetic vector potential is then
\begin{tbequation}
    \frac{\partial\AAA}{\partial t} = \frac{c}{q n_{\rm e}}\nab p_{\rm e}
\end{tbequation}
where $n_{\rm e}$ and $p_{\rm e}$ are the electron number density and pressure,
$c$ the speed of light, and $q$ the electron charge.

\paragraph*{The Durrive battery.}
Massive stars are surrounded by a region of ionized gas.
\citet{Durrive_Langer_2015} proposed that an electromotive force should
be created by the surplus momentum transferred to the electron after
the ionization of an atom.
Then, the uncurled induction equation for zero initial magnetic field becomes:
\begin{tbequation}
    \frac{\partial\AAA}{\partial t} = \frac{c}{q n_{\rm e}}\nab p_{\rm e}
    -\frac{c}{qn_{\rm e}}\dot{\bf{p}}_{\rm e}
    \label{eq:durrive}
\end{tbequation}
where $\dot{\bf{p_e}}$ is the rate of momentum transfer to the electrons
and \Eq{eq:durrive} also includes the Biermann battery.

In the early Universe, the Biermann battery can appear 
from local fluctuations in the sound speed right after recombination
\citep{naoz_narayan} and later,
around rippled
shocks, while both battery mechanisms should operate around ionization
fronts \citep{Subramanian_1994,Kulsrud+97,Gnedin2000}.
\cite{Garaldi2021} performed cosmological simulations testing, among
other scenarios, the efficiency of the Biermann and Durrive battery terms
through cosmic time. They found that, although the two batteries behave
similarly, the Durrive term produces systematically weaker magnetic
fields by approximately three orders of magnitude.
\end{textbox}

\subsubsection{Small-scale dynamos as a seed for the large-scale dynamo}

All dynamos require seed magnetic fields---even small-scale ones.
However, a small-scale dynamo grows much faster than a large-scale one.
It would therefore be able to produce equipartition strength magnetic
fields from much weaker seeds.
The idea has been discussed by \cite{Beck+94}.

Another related idea is to produce galactic seed magnetic fields
in the first stars.
The simplest form of this idea is that stars could pick up a Biermann
seed, which would then be amplified though a stellar dynamo and get
ejected with a supernova explosion at the end of the star's life.
This scenario has been explored in many cosmological simulations
\citep{beck2013,Katz2019,Vazza2017,MartinAlvarez2021}, and shows that
it can, in fact, magnetize galaxies very efficiently. However, these
simulations use unrealistically high values for the supernova-injected
magnetic field \citep{EN2022}.
Those fields could then well be large-scale ones, i.e., on the scale of
stars, but those would grow on an even shorter timescale, because stars
are much smaller than the envisaged turbulent eddies in the interstellar
medium.
Young stars and also their surrounding accretion disks can host powerful
dynamos that also drive magnetized winds; see the estimates
in \cite{Bra00} and corresponding mean-field simulations by
\cite{vonRekowski+03}.
Those winds could magnetize the surrounding interstellar medium and could
well produce much more efficient seeds for the galactic dynamo than any
battery mechanism.
The wind-based injection model would be a viable alternative to the
uncertain supernova seeding often used in cosmological simulations.

\subsubsection{Battery versus plasma instabilities}

When the electron distribution is anisotropic and the magnetic field
is not too weak, the Weibel instability \citep{weibel1959} can amplify the magnetic field,
Typically, the Weibel instability generates very small-scale fields.
Nevertheless, it could play a role in an intermediate regime when the
Biermann battery has generated a sufficiently strong magnetic field.
This is also in an agreement with recent laser plasma
experiments that have accessed a regime relevant to
astrophysical dynamos \citep{Schoeffler+16}.

\begin{marginnote}[]
\entry{Weibel instability}{Occurs in a nearly homogeneous plasma
when there is an anisotropy in velocity space}
\end{marginnote}

\subsubsection{Primordial seed magnetic fields}
\label{subs:bigbangturb}

In the early Universe, inflation and phase transitions, such as the
decoupling of the weak force and the electromagnetic force or the formation
of hadrons from quarks, may have produced hydromagnetic turbulence
\citep{Widrow02}.
Owing to the lack of further energy input, any magnetic field generated
at that time would be slowly decaying.
The dilution of the magnetic field due to the expansion of the Universe
is always scaled out by talking about the comoving magnetic field, which
is $\tilde{\BB}=a^2\BB$, where $a$ is the scale factor of the Universe.
When time is being replaced by conformal time, $\tilde{t}=\int\dd t/a(t)$,
the MHD equations, during the radiative era, have their usual
form without expansion factors \citep{BEO96}.
Hereafter, the tildes are therefore dropped.

Not much is known about the strength of the comoving magnetic field today.
There are only constraints.
Upper limits can be derived from Big Bang nucleosynthesis (BBN)
constraints; see \cite{Grasso+Rubinstein95} and \cite{Kahniashvili+22}
for recent work taking into account the decay of the magnetic field
between the moment of generation and the time of BBN.
A lower limit on the present-day magnetic field strength has been proposed
on the grounds that magnetic fields would prevent the reconnection
of pair-created electrons and positrons when the TeV photons from
powerful blazars interact with the extragalactic background light
\citep[e.g.,][]{neronov}.
Note, however, that the validity of this technique
may have a systematic uncertainty in that plasma instabilities could
potentially also provide an explanation for the non-observation of GeV
halos \citep{Broderick+12}.

Simulations by \cite{Sironi14}, confirm that plasma instabilities do
indeed operate, but they only account for about 90\% of the loss of
GeV photons and the suppression of the remaining 10\% would still need
to be explained by the presence of magnetic fields.
Similar conclusions were reached by \cite{AlvesBatista+19}, who performed
detailed simulations for individual blazars.

\begin{figure}[t]
\includegraphics[width=\textwidth]{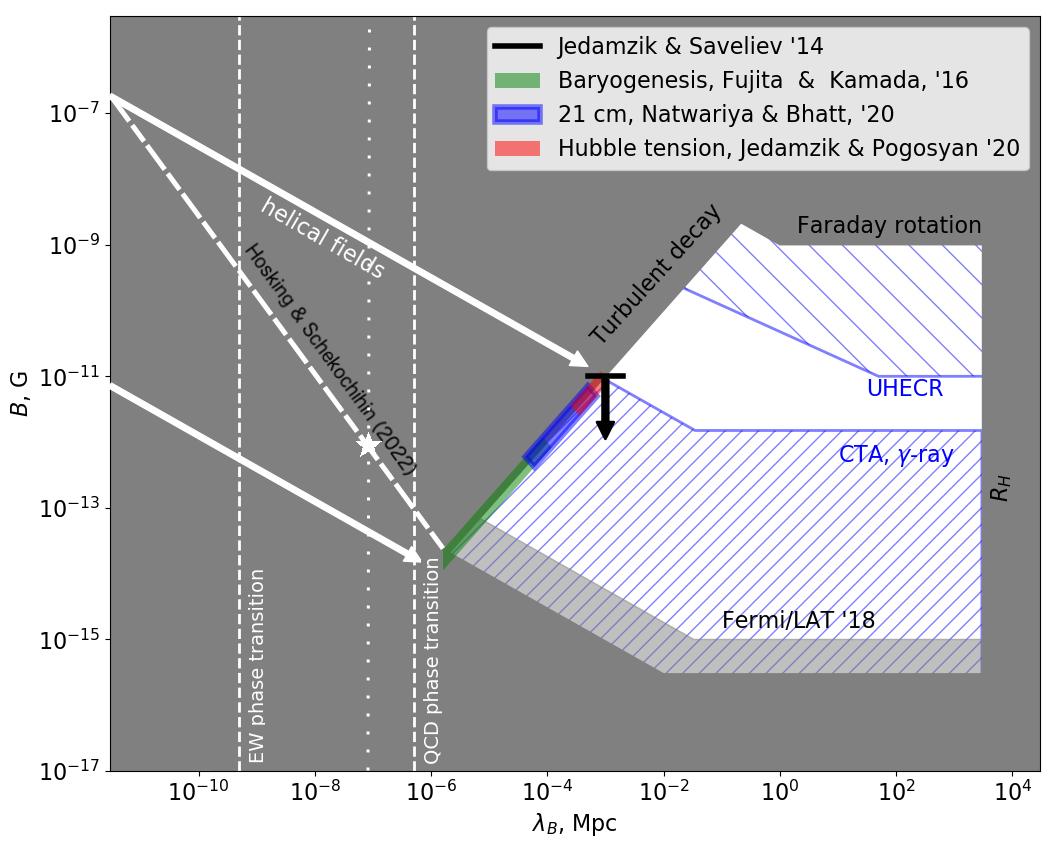}
\caption{
Summary of lower and upper magnetic field limits as a function of
correlation length.
The white solid lines describe the decay of a helical magnetic field
($\BB^2\sim t^{-2/3}$) along with the increase of its typical length scale
($\lambda_B\sim t^{2/3}$), so that $\BB^2\lambda_B=\const$.
Only the narrowly hashed region indicates a few permissible strengths.
Courtesy of \cite{Korochkin+21}, in which we have added the prediction
of \citet{Hosking_Schekochihin_2022} $B\propto\lambda_{B}^{-5/4}$.
The asterisk shows the scale where \citet{Hosking_Schekochihin_2022}
stop the line in their work, since they assumed that the relevant time
scale is determined by magnetic reconnection and not by the Alfv\'en time.
}\label{IGMF_hints}
\end{figure}

The lower limits derived by \cite{neronov}, which become less stringent
for larger length scales, provide an exciting motivation for primordial
magnetogenesis scenarios.
At the present time, those primordial magnetic fields
may have strengths in the range
of $10^{-16}\G$ to $10^{-9}\G$ \citep[see][for a review]{subramanian2016},
and could act as seed magnetic fields for any subsequent dynamo processes
-- once sufficient kinetic energy becomes available.
These seeds, if confirmed, would not only be stronger than those from
batteries, but they would also be present in the voids.
\Fig{IGMF_hints} shows the expected magnetic field ranges as a function
of the typical scale $\lambda_B$ of the magnetic field.
For a nonhelical magnetic field, there are still magnetic
helicity fluctuations.
They constrain the decay such that the correlation integral of the local
magnetic helicity is conserved \citep{Hosking+Schekochihin21}.
This leads to a decay with $\BB^2\sim t^{-10/9}$ and
$\lambda_B\sim t^{4/9}$, so that $\BB^4\lambda_B^5=\const$
\citep{Hosking_Schekochihin_2022}.

The first stars are expected to form about $10^{8}\yr$ after the Big Bang,
marking the beginning of the reionization epoch.
After that, galaxies start growing through continuous gas accretion
and mergers \citep{Dayal_ferrara2018}.
Since dynamo action is fastest at small length scales, the magnetic field
generation during the formation of the first collapsing structures is
potentially important and may have produced a stronger seed magnetic
field for the subsequent global galactic dynamo.
Strong magnetic fields may also affect galaxy and large-scale structure
formation of the Universe \citep{Kahniashvili+13}.

\subsubsection{Primordial fields during structure formation}
\label{sec:cosmosims}

Modern numerical simulations of cosmological large-structure
formation are taking into account the evolution of the magnetic field,
seeded by the various mechanisms outlined above.
A central question in these studies is whether the topology and strength
of these primordial fields leave measurable signatures on the cluster or
galaxy structures.

\begin{figure}[t]
\includegraphics[width=\textwidth]{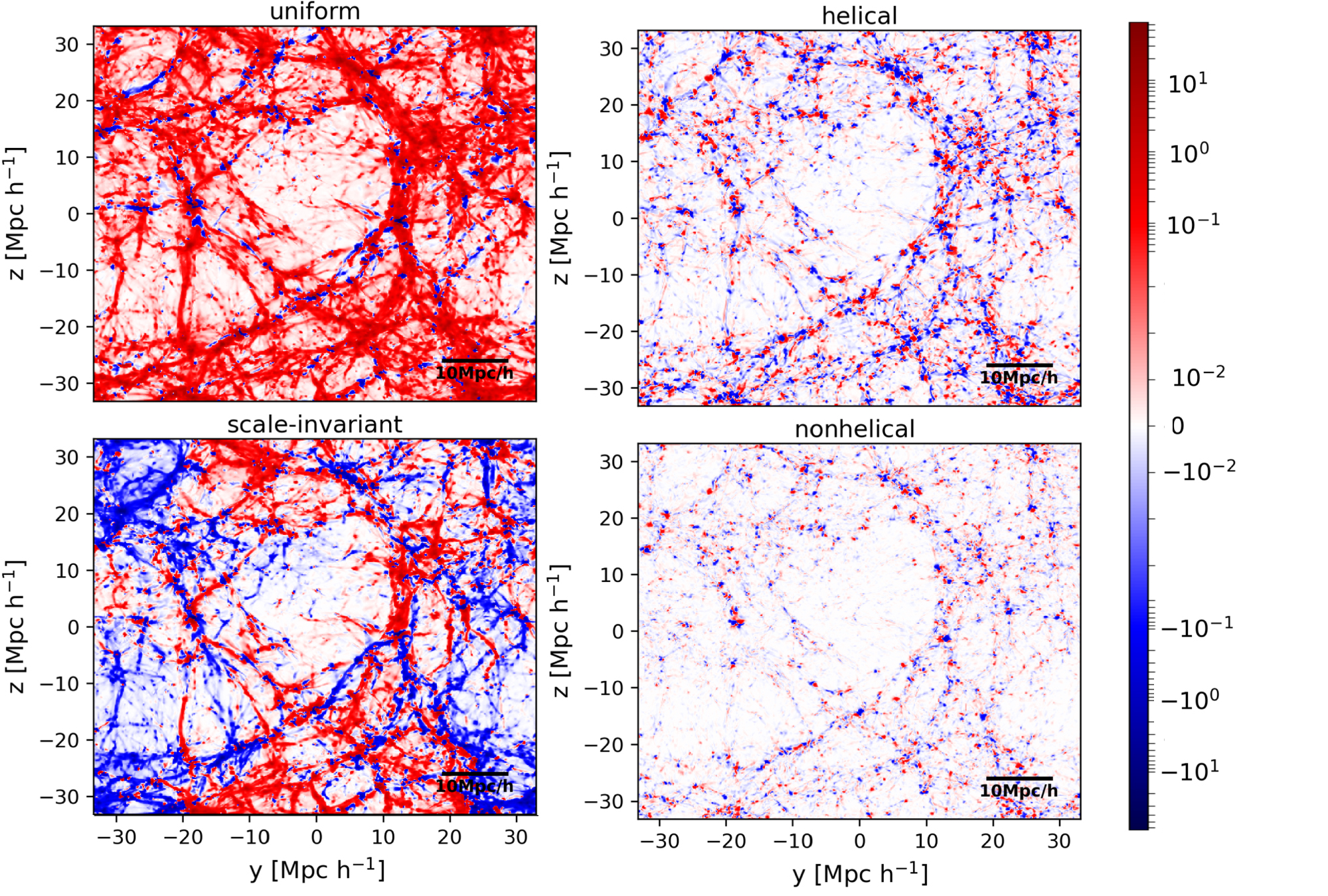}
\caption{
Faraday rotation maps from four cosmological simulations of
\cite{Mtchedlidze+22} using inflationary uniform and scale-invariant
fields (upper and lower left), as well as phase-transitional helical
and nonhelical initial fields (upper and lower right).
Courtesy of \cite{Mtchedlidze+22}.
}\label{Salome}
\end{figure}

\citet{Vazza2017} performed a comprehensive suite of cosmological
simulations using different magnetogenesis mechanisms: a uniform
seed, meant to simulate the magnetic field created by inflation,
a seed that follows the distribution of density perturbations to
approximate the magnetic field generation by a Biermann battery,
a seed that approximates the turbulent dynamo amplification, and an
astrophysical seed that simulates the injection of magnetic fields by
stellar sources.
They find that, at $z=0$, all mechanisms agree on the cluster
magnetization (which they were designed to reproduce).
However, there are large differences in the magnetic field structure
both on galaxy scales and in the voids.
Recently, \citet{Mtchedlidze+22} explored a more diverse set of
primordial magnetic fields, including uniform and
scale-invariant inflationary fields, as well as helical and
non-helical fields from the radiation-dominated epoch.
They also reported that the final magnetic field distribution retained
a memory of the initial seed.
This can be seen from \Fig{Salome}, where we show maps of Faraday
rotation at the present time.
The simulations started at a redshift of $z=50$ with
the four initial conditions discussed above.

The above works use a uniform spatial resolution, which offers the
advantage of an unbiased view of cosmological magnetic field evolution.
However, models with adaptive resolution can give a more detailed view of
the magnetic field on galaxy scales, while following their cosmological
history.
One recent example is the work of \citet{Garaldi2021}, who explored the evolution
of cosmological volumes and zoom-ins using four different mechanisms for
magnetic field generation: primordial, Biermann battery, Durrive battery,
and stellar seeds.
They report, contrary to the findings of the uniform-resolution,
large-volume works mentioned above, that the initial conditions
are forgotten by redshift $z\sim2$.
However, none of their initial conditions contained magnetic helicity,
which should not have decayed.

\citet{Marinacci2016} and \citet{MartinAlvarez2020} focused on the
effects of cosmological magnetic fields on galaxy formation.
They found that fewer, smaller galaxies form for stronger primordial
fields.
As mentioned in \Sec{sec:history}, \citet{MartinAlvarez2021},
traced the evolution of the primordial field and the field injected
by stellar sources separately (see \Fig{fig:martinalvarez}).
They achieved this by adding two tracer induction equations to the code,
one for each seed.
These induction equations are not connected to the gas evolution, but
only follow the evolution of the two fields as it would be if they were
independent of each other.
They found that their evolved galaxies contain a mixture of both:
metal-poor gas at the galaxy's outskirts containing mostly primordial
fields with large-scale coherence and supernova (SN)-enriched gas containing mostly
fields of stellar origin with small-scales coherence.
The cold, star-forming gas contains a mixture of the two.
However, in agreement with the \citet{Garaldi2021} result, the origin
of the galactic magnetic field becomes practically indistinguishable
very early on without the tracers.
All these simulations result in microgauss magnetic fields, but their
length scales are typically too small to explain the fields seen in
actual galaxies.

The results of these comprehensive simulations
point to a complex picture in which various seeding
mechanisms combine to give the initial and boundary conditions for
dynamos on different scales and different epochs.
They also point to cluster scales -- rather than galaxy
scales for an answer regarding the origin of cosmic magnetic fields.

\subsubsection{Possible importance of cluster mergers}

Mergers of galaxy clusters could amplify large-scale magnetic fields
quickly to near-equipartition strengths.
The merger itself could stretch a pre-existing field and amplify it in
conjunction with the existing (possibly helical) background turbulence.
One could then think of this as some kind of $\alpha\Omega$ dynamo,
where the $\Omega$ effect is associated with the large-scale shear
generated during the merger.
Such simulations were produced by \cite{Roettiger+99}.

The study of the relevance of cluster mergers to dynamos has not been
followed up in recent years.
In the meantime, there have been many relevant advances in dynamo theory
in connection with time dependence of the flow and in the context of
measuring field transport coefficients.
In view of these advances, this approach might deserve more detailed
follow-up studies in the future.
However, there are similarities with recent studies of gravitational
collapse dynamos that will be discussed next.

\subsection{Dynamos from gravitational collapse and other instabilities}
\label{sec:grav_collapse}

By the time the first gravitationally bound structures (stars or galaxies)
formed, any primordial turbulent velocities from the processes we mention
in \Sec{subs:bigbangturb} had already decayed.
However, the assembly into these first structures certainly generated
large amounts of turbulent kinetic energy, which could have triggered
dynamo action.

Several numerical works show that the formation of the first stars is
ideal for amplifying nG \citep{Sur+10,Sur+12,Federrath+11b},
or even just $10^{-20}\G$ \citep{Schober+12} fields to equipartition
values through a small-scale dynamo.
Gravitational compression can amplify the field further, although in the
presence of turbulence, the resulting dependence of the magnetic field
on the density is weaker than the prediction from ideal flux freezing
\citep{Sur+12}.
While flux freezing predicts $|\BB|\propto\rho^{2/3}$, the theory
of \citet{XL20}, which includes the turbulent dynamo, predicts that
$|\BB|/\rho^{2/3} \propto \rho^{2/57-1/6}\approx\rho^{-0.13}$,
where the scaling of the small-scale dynamo enters as an assumption.
This remarkable agreement with the simulation results of \cite{Sur+12}
is taken to be suggestive of the
importance of reconnection diffusion and the breakdown of flux-freezing
\citep{XL20}, which makes compressional amplification less efficient.

However, the results obtained so far still leave some questions unanswered.
For example, is it possible to explain the slow-down of
compressional field amplification even in the absence of
dynamo action?
Although this may be an academic question, it could be answered by
performing collapse simulations in two dimensions, when dynamo action
is impossible.
Also, it would be interesting to see the early magnetic field growth
starting from a much smaller initial magnetic field.
In fact, the theory of \cite{XL20} does not really address
when the dynamo is excited, but focusses on the discussion of the
nonlinear regime of a supercritical dynamo.

\subsubsection{Nature of collapse dynamos}

An important tool for characterizing dynamo action in time-dependent flows
such as decaying turbulence or gravitational collapse, is to compare
the work done against the Lorentz force with the Joule dissipation rate,
and to look at different contributions to the Lorentz work term.
These work and dissipation terms emerge when deriving the evolution
equation for the magnetic energy density.
Taking the dot product of \Eq{InductEq} with $\BB$, averaging, and
ignoring surface terms, we obtain
\begin{equation}
\frac{\dd}{\dd t}\bra{\BB^2/2\mu_0}=
\bra{\JJ\cdot(\UU\times\BB)}-\eta\mu_0\bra{\JJ^2}.
\label{dB2dt}
\end{equation}
Using $\JJ\cdot(\UU\times\BB)=-\UU\cdot(\JJ\times\BB)$,
one can write the first term on the right-hand side of \Eq{dB2dt}
as the work against the Lorentz force.
Two further refinements can then be employed \citep{BN22}.
First, one can decompose
$(\UU\times\BB)_i=-U_j\partial_j A_i+U_j\partial_i A_j$
to write $W_{\rm L}=W_{\rm L}^{\rm2D}+W_{\rm L}^{\rm3D}$,
where $W_{\rm L}^{\rm2D}=\bra{J_i U_j\partial_j A_i}$ and
$W_{\rm L}^{\rm3D}=-\bra{J_i U_j\partial_i A_j}$.
The second term, $W_{\rm L}^{\rm3D}$, vanishes for 2-D
magnetic fields oriented in the plane and describes therefore the work
term associated with 2-D compression, stretching,
and bending, such as in \Eq{InductEq2D}.
Second, one can decompose $\JJ\times\BB$ into contributions from
the magnetic pressure force, the tension force, and curvature force.
The corresponding work terms are then referred to as $W_{\rm L}^{\rm c}$
(for compression), $W_{\rm L}^\|$ (for tension force, i.e., along
the field), and $W_{\rm L}^\perp$ (for the curvature force, i.e.,
perpendicular to the field).

\begin{figure*}\begin{center}
\includegraphics[width=\textwidth]{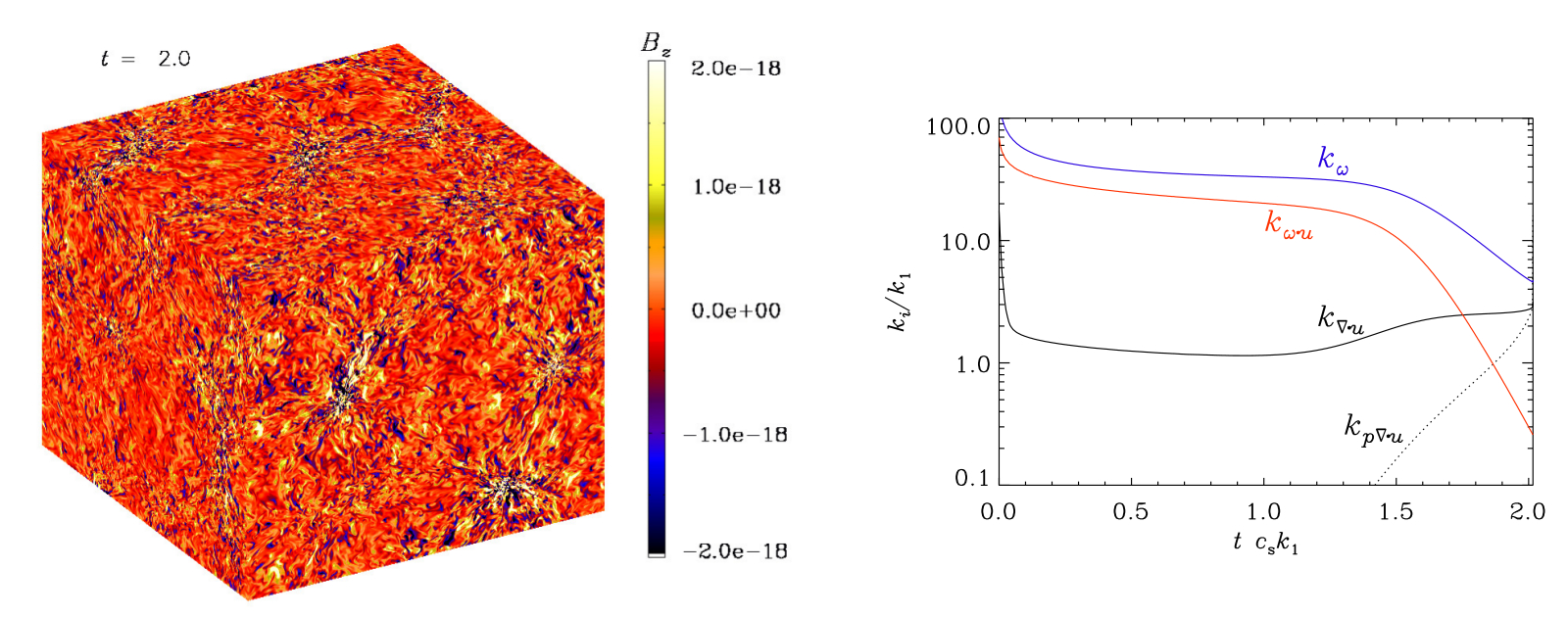}
\end{center}\caption[]{
Left: Visualization of the magnetic field.
Right: Characteristic wavenumbers $k_\omega$ (blue),
$k_{\oo\cdot\uu}$ (red), $k_{\,\nab\cdot\uu}$ (solid black),
and $k_{p\nab\cdot\uu}$ (dotted black).
}\label{bb3_0020}\end{figure*}

To determine the reality and nature of dynamo action during a turbulent
self-gravitational collapse more carefully, \cite{BN22} computed
the aforementioned terms that enter the magnetic energy balance.
The basic conclusion is that there is indeed dynamo action during the
early phase of the collapse while the initial turbulence is slowly
decaying, but that dynamo action diminishes when the flow becomes
dominated by 3-D compression toward the various collapsing
potential minima, where only the irrotational flow component gains in
strength, which, however, does not (or not much) contribute to dynamo
action in their simulations.
In \Fig{bb3_0020} we visualize the collapsing magnetic
field from \cite{BN22},
the diminishing of the vorticity, expressed here as a wavenumber
$k_{\oo}=\orms/\urms$, the gain of compressive motions, here expressed through
$k_{\,\nab\cdot\uu}=(\nab\cdot\uu)_{\rm rms}/\urms$, along with several
other quantities, $k_{p\nab\cdot\uu}=-\bra{p\nab\cdot\uu}/p_0\urms$
and $k_{\oo\cdot\uu}=|\bra{\oo\cdot\uu}|/\urms^2$, characterizing the
work done by compression and the amount of kinetic helicity, respectively.
A potential problem with the simulations of \cite{BN22} is the relatively
short collapse time compared with the turnover time of the turbulence.

\subsubsection{Magneto-buoyancy and magneto-rotational instabilities}

These instabilities can drive turbulence and may play important roles
in parts of the galaxy.
\begin{marginnote}
\entry{Magneto-rotational instability}{arises in a magnetized, rotating disk when the angular velocity decreases with distance.}
\end{marginnote}
Buoyancy may be driven by cosmic rays inflating flux tubes and are
thought to speed up the dynamo \citep{Parker92,Hanasz2013}.
The magneto-rotational instability \citep{BH1991} can drive turbulence from the
kinetic energy in the shear.
It can also play a role in the outer parts of the galaxy were supernova
driving is less efficient \citep{Piontek+Ostriker07}.

\section{GALACTIC MEAN-FIELD DYNAMOS}
\label{sec:mean-field_models}

\subsection{Global magnetic field structure}

One of the strongest existing tests for dynamo theories is the predicted 
structure of the large-scale magnetic field contrasted to observations.
In the next sections we outline the predictions from different models.

\subsubsection{Early analytic approaches}

The idea that the large-scale magnetic field of galaxies could be
explained through an $\alpha\Omega$ dynamo was formulated early on
\citep{VR71,Par71}, just after the first successful mean-field models
were proposed for the Sun and Earth.
An important early result was the finding that the most preferred
magnetic field mode in flat geometries like galaxies is quadrupolar,
i.e., the toroidal field is even about the midplane.
[Here and elsewhere, quadrupolar means not just a quadrupole, but all
modes of even symmetry about the midplane \citep{KR80}.]

In view of many early claimed discoveries of BSS fields \citep[for
a review, see][]{SFW86}, an important question in those early days
concerned the possibility of preferred nonaxisymmetric magnetic fields.
Such modes were never found.
However, when the assumption of what is known as the pure $\alpha\Omega$
approximation was made, i.e., the toroidal field is only generated
by differential rotation ($\Omega$ effect) and the $\alpha$ effect is
\begin{marginnote}[]
\entry{ASS and BSS fields}{defined as $m=0$ and $m=1$ magnetic 
fields and vary with azimuth $\phi$ like $e^{\ii m\phi}$.}
\end{marginnote}
neglected in the generation of the toroidal magnetic field,
nonaxisymmetric modes where found to be excited, although the growth rates
of the corresponding ASS fields were always larger; see \Sec{sec:history}.
This approximation turned out to be not permissible, when the
magnetic field is nonaxisymmetric.

\subsubsection{Boundary conditions}

Standard dynamo problems are usually formulated with vacuum
boundary conditions, i.e., the magnetic field is current-free
and extends to infinity outside the domain \citep{KR80}.
However, such boundary conditions can only be formulated for
spheres or ellipsoids, but not for cylinders, for example.
\cite{Stix75} employed ellipsoidal coordinates and obtained an
axisymmetric solution.
Contrary to \cite{Par71}, he found that oscillatory solutions
occur only at substantially larger dynamo numbers.
\begin{marginnote}
\entry{Dynamo number}{Defined as $D=R_\omega~R_\alpha$, where $R_\omega$
and $R_\alpha$ are mean-field magnetic Reynolds numbers associated with
differential rotation and $\alpha$-effect, respectively.}
\end{marginnote}
Unfortunately, the implementation of ellipsoidal coordinates in a
numerical code is rather cumbersome.
This led to the approach of embedding the galaxy in a sufficiently
large poorly conducting halo, which itself is then contained either
in a cylinder with perfectly conducting boundaries \citep{Elstner+90},
or in a sphere with vacuum boundaries \citep{BTK90}.
These two alternatives are rather different from each other, but the hope
is that these boundaries are far enough away from the physical boundaries
that these differences are without consequence.

\subsection{Dynamo models for specific galaxies}

Various attempts have been made to produce dynamo models of individual
galaxies.
One such example is M31, i.e., the Andromeda galaxy.
Its magnetic field is often described as a ring field.
It is also often regarded as an analog of the Milky Way.
Corresponding models have been presented by \cite{Poezd+93} using
nonlinear $\alpha$ quenching.
An important challenge here is to reproduce the right pitch angle
of the magnetic field and its radial dependence; see \cite{Shukurov00}
and \cite{Fletcher+04} for detailed discussions.

Another interesting case is M81, whose magnetic field is possibly
predominantly nonaxisymmetric.
This was difficult to explain.
\cite{Moss+93} showed however that such a field could result from an
initial magnetic field that might have survived for long enough times,
at least in the outer parts of the galaxy.
Yet another very different case is NGC~6946, whose field may consist
of structures usually termed magnetic arms. Magnetic arms are often
interlaced with the stellar arms, but can also be phase-shifted relative
to them; see \cite{Shukurov98} and \cite{Chamandy+15} for a more detailed
discussion, and \cite{Beck_2019} for recent updates.

Finally, we mention the magnetic fields in the halos of the edge-on
galaxies NGC~891 and NGC~4631.
\cite{BDMSST93} and \cite{Elstner+95} found that the observed polarization
vectors could only be reproduced when there is a strong enough outflow.
We return to this in \Sec{SynchrotronFromModels}.

\subsection{Galactic models with magnetic helicity flux}

In \Sec{sec:helicityflux} we have discussed the potential importance of
magnetic helicity fluxes.
It is often believed that they would be required to explain strong
magnetic fields in galaxies.
Here, we demonstrate the effect of magnetic helicity fluxes in specific
models.
\cite{SSSB06} have presented nonlinear models in a one-dimensional
geometry using the magnetic helicity flux associated with a galactic
fountain flow.
The main nonlinearity was here given through the dynamical quenching
formalism with advective magnetic helicity fluxes included,
similarly to what was discussed in \Sec{sec:helicityflux}.
The authors found that a magnetic helicity flux does indeed lead to
larger magnetic field amplitudes provided the magnetic helicity flux
is strong enough.
In their model, the helicity flux was accomplished through a galactic
fountain flow with a speed of at least $300\m\s^{-1}$.
More detailed studies have been performed by \cite{Prasad+16},
who also included advective and diffusive magnetic helicity fluxes.

\begin{figure*}\begin{center}
\includegraphics[width=\textwidth]{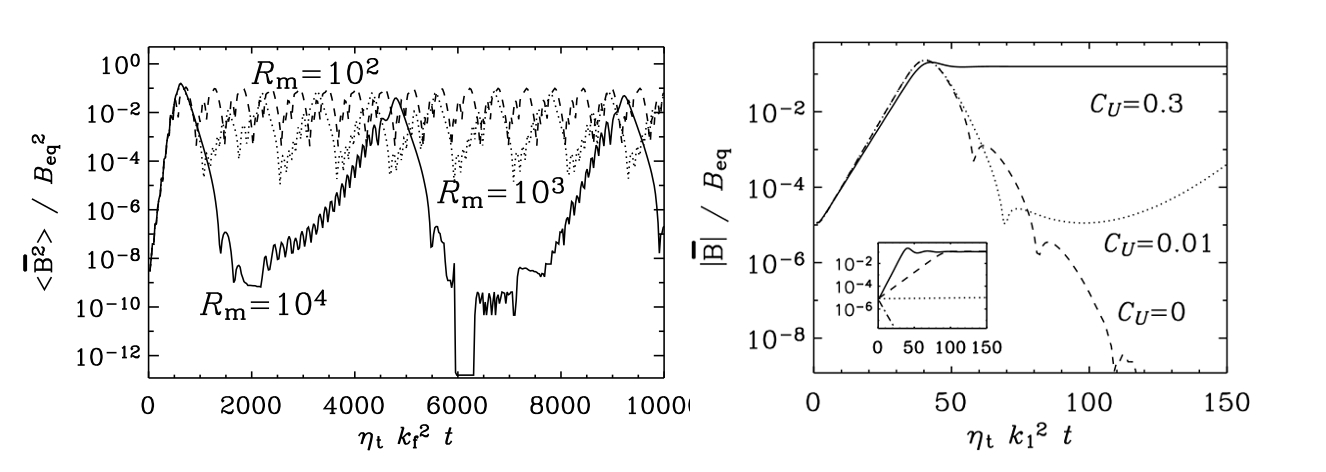}
\end{center}\caption[]{
Left: Evolution of the energy of the mean magnetic field in a model with
a shear-induced magnetic helicity flux for different values of $\Rm$
(here denoted by $R_{\rm m}$).
Adapted from \cite{Bra+05}.
Right: Magnetic field evolution in models with advective magnetic helicity
fluxes for $\Rm=10^5$ and different values strength of advection.
The advection velocity is characterized by the parameters $C_U$.
Courtesy of \cite{SSSB06}.
}\label{Hl232_f1}\end{figure*}

There is some uncertainty regarding the main contributors to the magnetic
helicity flux; see the more recent study by \cite{Vishniac+Shapovalov14}.
In addition to advection, shear could modify the turbulent correlations
in such a way as to transport magnetic helicity efficiently outward.
First proposed by \cite{Vishniac+Cho01}, this can lead to episodic
magnetic field amplification, especially as $\Rm$ is increased; see
\cite{Bra+05} and the left-hand side of \Fig{Hl232_f1}.
On the right-hand side of \Fig{Hl232_f1}, we reproduce the simulation
result of \cite{SSSB06} with an advective magnetic helicity flux.
In the models with insufficient advective flux, the magnetic energy
decreases to very small values.
The earlier simulations by \cite{Bra+05}, their Figure~7, with somewhat
smaller values of the magnetic Reynolds number showed that the magnetic
field can recover after some time, but then, again, it begins to fall off,
just like what is seen in \Fig{Hl232_f1}.
However, one may want to remain sceptical about whether these fluxes really do
alleviate the catastrophic quenching, because so far this has been 
seen only in mean-field models and not yet in actual turbulence simulations.

\subsection{Galactic rotation measure signature}

In \Sec{sec:history} we mentioned the historical importance of
RM studies for distinguishing between an ASS field, characteristic of dynamo
models, and a BSS field, characteristic of a wound-up primordial fields.
The subsequent findings of RM studies indicate that galactic magnetic field
evolution might be more complex than this simple dichotomy.

At the time of the review of \cite{SFW86}, most galaxies were
thought to be of BSS type; the authors listed seven out of 11 galaxies
as having a BSS field.
However, more accurate subsequent surveys confirmed a predominantly BSS
type structure for only M81 \citep{KBH89}.
In a later review, \cite{Beck+96} listed the field structures for 33
galaxies.
The picture became more complicated, with four examples primarily of
ASS fields (albeit two where marked as uncertain).
The dominance of ASS over BSS continues to persist even today.
For M33, \citet{Tabatabaei+2008} found an axisymmetric field in the
inner regions and a superposition of axisymmetric and bisymmetric fields
in the outer regions.
\citet{Beck_2015} observed a weak ($0.5\uG$) axisymmetric field in
IC~342, and \citet{Beck_2020} found a dominating ASS field in M31
combined with a six-times weaker BSS component.

\begin{figure}
    \centering
    \includegraphics[width=\textwidth]{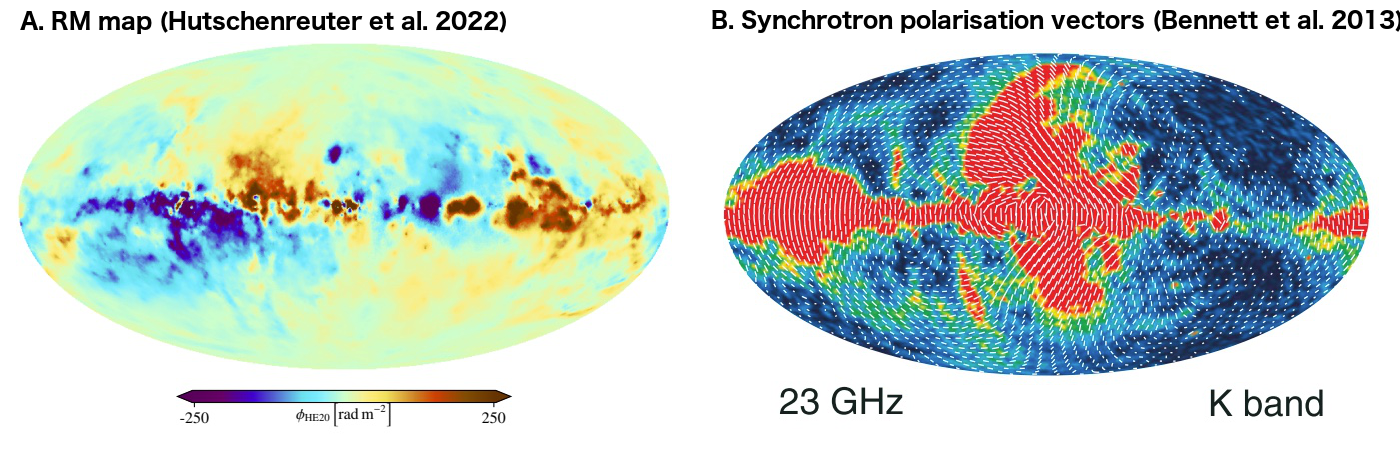}
    \caption{Full-sky view of the Galactic magnetic field in (A) RM of
    extragalactic sources
    \citep[][where the range of RMs is saturated at $\rm \lvert RM\rvert=250~rad/m^{2}$]{Hutschenreuter2022} %
    and (B) synchrotron
    emission \citep{Bennett2013}.}
    \label{fig:MWobservations}
\end{figure}

Classifying the Milky Way's magnetic field structure is much harder.
The existing parametric models of the Galactic magnetic field
are largely based on full-sky RM maps of extragalactic sources
\citep[e.g.,][]{oppermann2015,Hutschenreuter2022} and synchrotron
emission maps of the Milky Way \citep[mainly from the Wilkinson
Microwave Anisotropy Probe; see][see also
\Fig{fig:MWobservations}]{Page2007,Bennett2013}.
The model of \citet{sun2008} assumes an axisymmetric spiral with a
reversal in the inner 5~kpc.  \citet{Jansson_Farrar_2012} include
magnetic spiral arms and an X-shaped field in the halo (see also
\Sec{SynchrotronFromModels} for the observational motivation).
\citet{Jaffe2010} also fit magnetic spiral arms to the disk data, also
including the random magnetic field component.
\citet{Terral_Ferriere_2017}, using analytic forms for the
3-D field, conclude that a bisymmetric ($m=1$) halo field
best fits the RM data.
However, \citet{West2020} found evidence for an axisymmetric ($m=0$)
quadrupolar magnetic field with a small net vertical component in RM.
A newer analysis \citep{Dickey+22} shows that a combination of an
axisymmetric and a bisymmetric mode, based on analytical galactic
dynamo models from \citet{Henriksen_2018}, best explains the large-scale
morphology of the Galactic RM data.
As the quality of the observational data improves, a complex picture
of galactic magnetic field morphology emerges, including that of the
Milky Way.

\subsection{Synchrotron emission from mean-field models}
\label{SynchrotronFromModels}

Synchrotron emission provides an important means of measuring the magnetic
field in galaxies and comparing models with simulations \citep[see][for
a review on bridging dynamo models and observations]{Beck_2019}.

Early attempts of computing the polarized synchrotron emission from
models were presented by \cite{DB90}, who computed the linearly
polarized emission from galactic mean-field models, which contained
both axisymmetric and nonaxisymmetric magnetic fields.
They confirmed the idea of distinguishing these modes by measuring
the RM along a ring around the galaxy.
Another example was the computation of linear polarization from the
magnetic field on both sides of the midplane in edge-on galaxies.
In galaxies seen edge-on, synchrotron
emission reveals X-shaped halo magnetic fields
\citep[e.g.,][]{Golla_hummel_1994,Tullmann2000,Krause2006,stein2019,krause2020}.
Although the 3-D morphology of these fields is unknown,
the X-shaped signature can be reproduced by dynamo models that include
an outflow \citep{BDMSST93,Elstner+95}.
The resulting magnetic fields were thought to have quadrupolar symmetry
also in the halo, but this now seems to be ruled out by new observations
\citep{NGC4631}.
An alternative would be a dynamo in the halo itself, which could produce
dominant dipole modes \citep{BDMSST92,Moss_sokoloff_2008}.

At long radio wavelengths, the synchrotron emission from even just a
uniform magnetic field suffers depolarization from the superposition of
Faraday-rotated contributions.
However, if the magnetic field is helical, the polarized intensity can
either enhance the depolarization if helicity and RM have opposite signs,
or it can cancel it if they have the same sign \citep{BS14,HF14}.
This leads to a correlation between polarized intensity and RM
\citep{Volegova+Stepanov09}, which has now been used by
\cite{West2020} to characterize the Galactic magnetic helicity.
Future observations with the Square Kilometre Array are expected to reveal
much more detailed information on magnetic helicity using a continuous
band of wavelengths \citep{Beck+15}.

The synchrotron intensity gives an indication about the magnetic field
strength.
It is proportional to the product of the density of relativistic cosmic
ray electrons and a power close to 2 of the local magnetic field component
perpendicular to the line of sight.
However, the relativistic cosmic ray electron density may itself depend
on the local magnetic energy density, because cosmic rays and magnetic
fields have supernova explosions as a common source of energy.
These arguments have been reviewed by \cite{Seta+Beck19}, who also make
comparisons with numerical simulations of cosmic ray confinement by a local
dynamo-generated magnetic field, similar to what was done earlier by
\cite{Snodin+06}.
\cite{Seta+Beck19} conclude that the commonly made assumption of an
equipartition between cosmic ray and magnetic energy densities is not
valid on scales smaller than at least $100\pc$.
They argue that ignoring the nonlinear dependence of the synchrotron
emission on the plane-of-the sky magnetic field component can lead to
an overestimation of the actual magnetic field by up to a factor of 1.5.

An interesting comparison between radio synchrotron and dust polarization
in emission can be found in \citet{Borlaff2021} for M51.
They find that the magnetic pitch angles of the two tracers differ, with
the dust polarization showing a more tightly wound spiral than the radio.
In light of this and forthcoming comparisons, predictions from dynamo
models should take into account the multi-phase nature of the ISM.

\subsection{$E$ and $B$ polarizations}

The linear polarization described by Stokes $Q$ and $U$ can also
be expressed in terms of the rotationally invariant parity-even $E$
polarization and the parity-odd $B$ polarization, as is commonly done
in cosmology \citep{Seljak+97,Kamionkowski+97}.
Here, the symbols $E$ and $B$ have nothing to do with electric and
magnetic fields, except that both can qualitatively be described as
gradient-like and curl-like fields.
It is important to stress, however, that $E$ and $B$ are only defined
on a 2-D surface.
Therefore, the parity-odd $B$ polarization has no immediate correspondence
with the helicity of the underlying magnetic field.

Mathematically, $E$ and $B$ are obtained as the real and imaginary parts of
a quantity $R(\theta,\phi)$ with
\begin{equation}
R\equiv E+\ii B=\sum_{\ell=2}^{N_\ell}\sum_{m=-\ell}^{\ell}
\tilde{R}_{\ell m} Y_{\ell m}(\theta,\phi).
\end{equation}
and $\tilde{R}_{\ell m}$ are coefficients that have been computed as
\begin{equation}
\tilde{R}_{\ell m}=\int_{4\pi}
(Q+\ii U)\,_2 Y_{\ell m}^\ast(\theta,\phi)\,
\sin\theta\,\dd\theta\,\dd\phi,
\label{EBfromQU}
\end{equation}
with $_2 Y_{\ell m}(\theta,\phi)$ being the spin-2 spherical harmonics
and the asterisk denoting the complex conjugate.

\begin{figure}[t]
\begin{center}
\includegraphics[width=\textwidth]{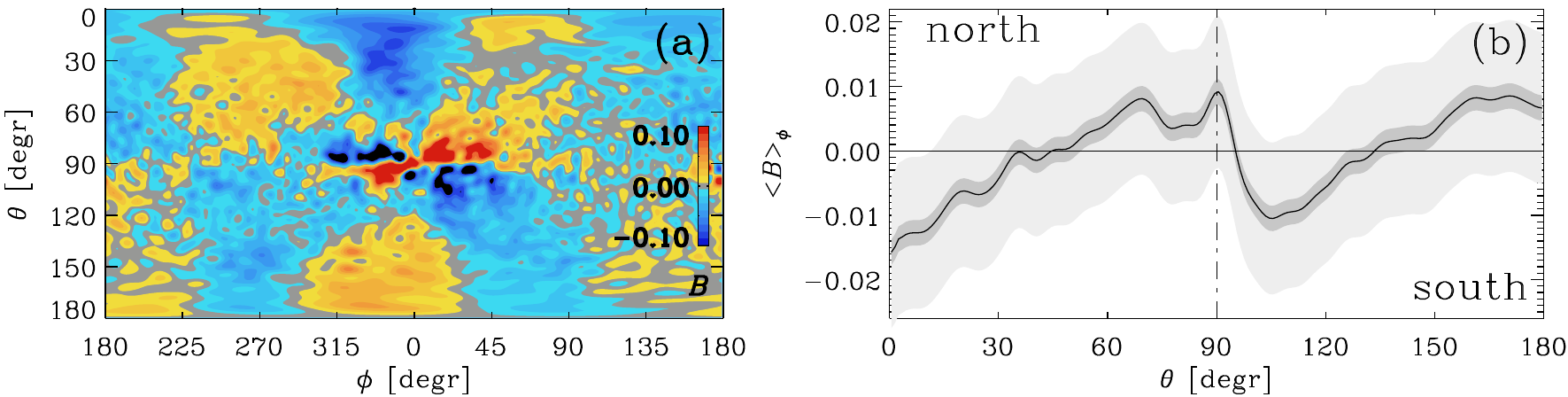}
\end{center}
\caption{Left: Galactic $B$ mode polarization.
Right: longitudinally averaged $B$ mode polarization.
Here, $\theta$ and $\phi$ are Galactic colatitude
($=90\degr-\mbox{latitude}$) and longitude.
}\label{RfromP_shift_last}
\end{figure}

\cite{BB20b} found that the $B$ polarization averaged over Galactic
longitude is very small owing to longitudinal cancelation, but there is
a small net hemispheric antisymmetry.
This is shown in \Fig{RfromP_shift_last}, where we plot the
Galactic $B$ mode polarization together with the
longitudinally averaged $B$ mode polarization.
It may be tempting to associate this hemispheric dependence with that
anticipated for the $\alpha$ effect, which is also a parity-odd quality
with hemispheric sign change.
However, the observed hemispheric antisymmetry
is actually explained by the spiral nature of the magnetic field.
Looking toward northern and southern galactic latitudes yields
mirror images of each other, which explains the observed hemispheric
antisymmetry of the mean $B$.

\section{TURBULENCE SIMULATIONS OF GALACTIC DYNAMOS}

\subsection{Physical parameters of the ISM}

The interstellar gas can be found in various phases, characterized by different temperatures, densities, and degrees of ionization.
The neutral, atomic gas is found in a cold ($50\K<T<100\K$) and a warm
($10^3\K<T<10^4\K$) phase, usually termed the cold and warm
neutral media (CNM and WNM, respectively).
The ionized gas is also found in a hot ($T\simeq10^6\K$) and a warm
($T\simeq10^4\K$) phase (HIM and WIM, respectively).
Finally, the densest and coldest ($10\K<T<20\K$) gas is mostly
molecular medium (MM), with a very low ionization fraction.

Dynamos can easily be excited in the interstellar medium, since
all phases (apart from the MM) are characterized by large values of
$\Rey$, $\Rm$, and $\Pm$, although these vary greatly between phases;
see \Tab{Tinterstellar}, with order-of-magnitude values taken from
\citet{Ferriere2020} and \citet{Draine_2011}.
This vast range of parameters also poses a challenge for accurately
modeling interstellar turbulence.
At large $\Pm$, the small-scale magnetic energy tends to dominate and the
dynamo returns much of the magnetic energy back into kinetic energy \citep{BR19}.
However, no effect on the large-scale dynamo has been reported as yet.

\begin{table}
\tabcolsep7.5pt
\caption{
Parameters of the interstellar medium. 
}
\label{Tinterstellar}
\begin{center}
\begin{tabular}{@{}c|c|c|c|c|c|c@{}}
\hline
Phase & HIM & WIM & WIM & CNM & WNM & MM  \\
\hline
$\Rey$ & $10^2$    & $10^7$    & $10^7$    & $10^{10}$ & $10^7$    & $10^7$   \\
$\Rm$  & $10^{23}$ & $10^{19}$ & $10^{18}$ & $10^{11}$ & $10^{18}$ & $10^4$   \\
$\Pm$  & $10^{21}$ & $10^{11}$ & $10^{11}$ & $10^{4}$  & $10^{11}$ & $10^{5}$ \\
\hline
\end{tabular}
\end{center}
\end{table}

Most of the numerical work on ISM turbulent dynamos so far has been
isothermal \citep[e.g.,][]{Scheko+04,Seta+2020}.
In an interesting extension, \citet{Seta+Federrath22} modelled a
small-scale dynamo in driven turbulence simulations of a two-phase ISM
and identified the processes responsible for vorticity generation in
each phase.
They found that the magnetic to turbulent kinetic energy ratio is lower
in the cold phase.
This will be discussed in more detail in \Sec{LocalDynamo}.

\subsection{Numerical approaches}
\label{NumericalApproaches}

Including magnetic fields in simulations of astrophysical systems is a
non-trivial task because the chosen discretization must obey the zero
divergence constraint.
In Eulerian codes, a commonly used approach is the Constrained Transport
(CT) scheme \citep{Evans_Hawley1988}, which ensures $\nabla\cdot\BB=0$
by defining the magnetic field components on cell faces.
However, no similar scheme is applicable to Lagrangian codes such
as Smoothed Particle Hydrodynamics (SPH), which 
rely on
divergence-cleaning algorithms
\citep[e.g.,][]{Brackbill_Barnes_1980,Powell1999,dedner_2002}.

\begin{table}
\tabcolsep7.5pt
\caption{
Overview of the numerical codes mentioned in this review,
indicated by the references.
}
\label{Tcodes}
\begin{center}
\begin{threeparttable}
\begin{tabular}{@{}l|l|l@{}}
\hline

Code name & Approach & Other properties \\
\hline
AREPO\tnote{1} & $\nab\cdot\BB$-clean & 
\begin{tabular}{@{}l@{}}
Finite volume, unstructured moving mesh 
\end{tabular} \\
\hline
ENZO\tnote{2}  & CT or $\nab\cdot\BB$-clean & 
\begin{tabular}{@{}l@{}}
AMR, Riemann, split and unsplit schemes 
\end{tabular}\\
\hline
FLASH\tnote{3} & CT or $\nab\cdot\BB$-clean & 
\begin{tabular}{@{}l@{}}
AMR, Riemann, split and unsplit schemes 
\end{tabular} \\
\hline
Gadget\tnote{4} & $\nab\cdot\BB$-clean & SPH \\
\hline
Nirvana\tnote{5} & CT & AMR, Godunov (Riemann) \\
\hline
{\sc Pencil Code}\tnote{6} & $\BB=\nab\times\AAA$ & centered finite diff., sixth order \\
\hline
RAMSES \tnote{7} & CT & AMR, Godunov (Riemann) \\
\hline
\end{tabular}
\begin{tablenotes}
\footnotesize
    \item[1] \citet{Pakmor2017,vdv2021,Whittingham2021}
    \item[2] \citet{Vazza2017}, \cite{Mtchedlidze+22}
    \item[3] \citet{Sur+10,Federrath+11b,Sur+12}
    \item[4] \citet{Steinwandel2019}
    \item[5] \citet{Gressel2008}
    \item[6] \citet{Bra01,Bra05,Bra10,Bran+10,Bra19,Gent2021,BN22}
    \item[7] \citet{Rieder_Teyssier_2016,Rieder_Teyssier_2017a,Rieder_Teyssier_2017b,Ntormousi+20}
  \end{tablenotes}
\end{threeparttable}
\end{center}
\end{table}

\begin{figure}
    \centering
    \includegraphics[width=\textwidth]{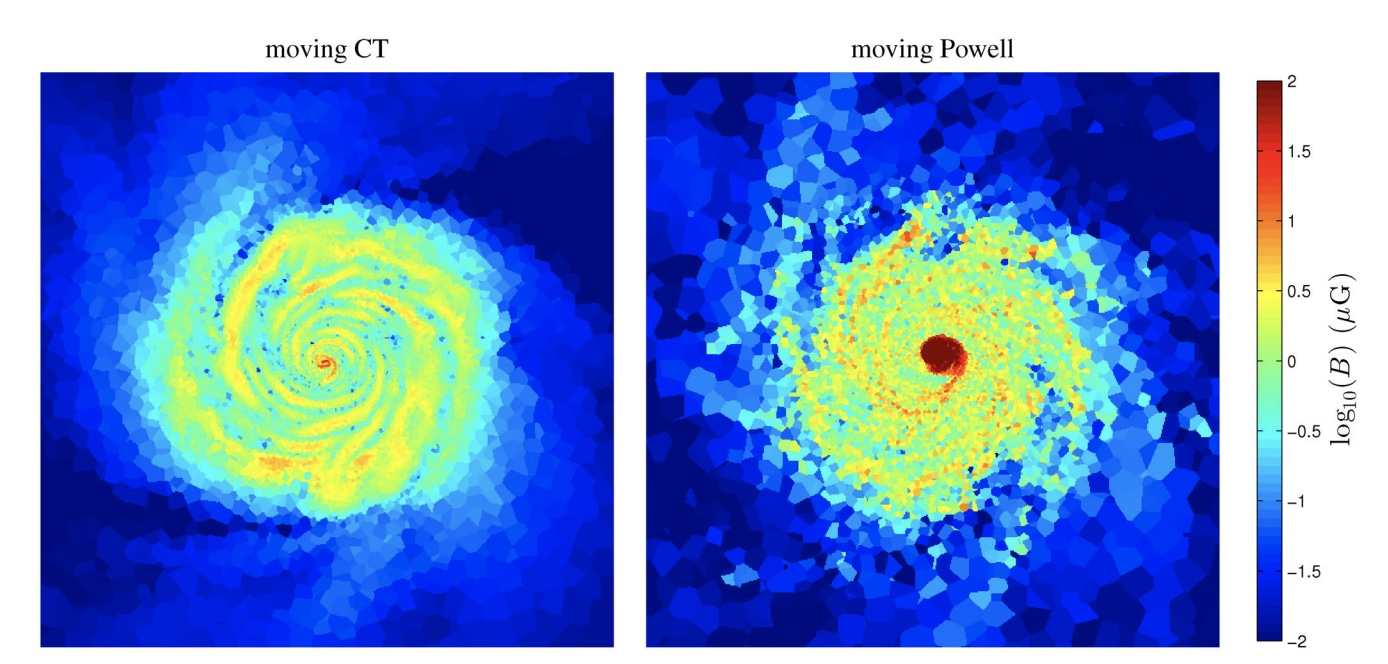}
    \caption{Figure~7 from \citet{Mocz2016}, where the same disk is evolved with CT and a Powell-like divergence cleaning scheme. The winding pattern of the magnetic field is better captured with CT, while an artifact is formed at the center with divergence cleaning.}
\label{divclean_mocz}
\end{figure}

Lagrangian codes are particularly well-suited for modeling galaxies and
cosmological volumes due to the natural adaptation of the resolution to
areas of interest.
They are also naturally Galilean-invariant.
However, their dependence on divergence cleaning poses a significant
drawback when modeling astrophysical dynamos.
This was clearly demonstrated by
\citet{Mocz2016}, who compared CT and divergence cleaning
approaches, both implemented on the moving-mesh code AREPO.
They found that divergence cleaning systematically creates artifacts
that mimic physical effects. Some of these artifacts are illustrated
in \Fig{divclean_mocz}.

Out of these effects, particularly notorious for dynamo studies
is an artificial increase of the magnetic energy when using scalar
divergence-cleaning schemes, as pointed out by \citet{Balsara_Kim_2004}
for supernova-driven turbulence.
Another very relevant example of an artifact caused by divergence cleaning
is the spontaneous production of magnetic helicity, which has been found
by \cite{BS2020}, who compared simulations that employed a divergence
cleaning algorithm with one that advances instead the magnetic vector
potential, $\AAA$, so that $\BB=\nab\times\AAA$ is always divergence-free.
They found that, for a helically driven flow in a periodic domain,
spurious net magnetic helicity is generated on dynamical timescales.
An interesting experiment by \citet{Tricco_2016} compared an SPH code
to the FLASH grid code, both using divergence cleaning, in simulations
of turbulent dynamos.
They found very good agreement between the codes, both in the growth
rates and the saturation level of the dynamo.
This suggests that potential problems with divergence cleaning
may not be severe.

Yet another method of dealing with the $\nab\cdot\BB=0$ constraint is
to employ the Euler or Clebsch potentials.
However, this method only works in the strictly ideal case
when the microphysical magnetic diffusivity vanishes; see
\Sec{NeedMagneticDiffusivity}.
As we have stressed in \Sec{NeedMagneticDiffusivity}, the addition of an almost negligibly
small diffusivity to the evolution equations for the Euler potentials
does not correspond to any physical magnetic diffusivity and leads to
wrong results where no dynamo is possible---even for flows that are
fast dynamos \citep{Bra10}.

\Tab{Tcodes} outlines some characteristics of the codes mentioned
in this review.
Next to each code we have mentioned the works cited in this review that use it.
The fact that many of them have to rely on divergence cleaning methods
is evidence of the difficulty in dealing with the divergence problem,
but also a sign to use caution when interpreting the results in the
context of dynamo action.

\subsection{Local dynamo simulations of galaxy portions}
\label{LocalDynamo}

Simulating the magnetic field evolution over an entire galactic disk is
another challenging task, due to the vast range of dynamical scales of
the problem and the large shearing velocities involved.
One approach to this challenge that can successfully capture many aspects
of the problem is simulating galaxy portions.
The local approach has been rather successful in the context of accretion
disks, where simulations have been performed in what is known as
shearing boxes.
This means that the radial boundary is ``shearing-periodic'', i.e., it
is periodic with respect to an azimuthal position that shifts in time
following the background shear flow.

Using a shearing box, \citet{Gressel2008} performed the first
simulation of a galactic dynamo, including supernova-induced turbulence.
It was similar to earlier multiphase simulations of supernova-driven
turbulence of \cite{Korpi+99}, where the magnetic Reynolds number was
still too low to permit dynamo action.
\cite{Gressel2008} found that the rotation frequency of the considered
galaxy portion is the dominant factor in determining the dynamo
efficiency, while the supernova rate did not significantly affect the
efficiency of the dynamo.
This finding suggests that the simulations were able to capture
large-scale dynamo action, but not small-scale dynamo action.
Interestingly, they also found no evidence of catastrophic quenching in
the range of $\Rm$ values explored by varying the rotation frequency
of the galaxy portion.
They hypothesize that this could be due to helicity fluxes.
In a subsequent paper, \cite{Gressel+13} speculated about various
quenching scenarios based on the magnetic field dependence, but that
was just for one value of the microphysical magnetic diffusivity.

Using the {\sc Pencil Code} and a similar setup, \citet{Gent2013}
showed that the mean and fluctuating fields have different growth rates,
indicating a co-existence of small- and large-scale dynamos.
Theoretically, however, the possibility of large-scale and small-scale
dynamos having different growth rates in one and the same system is not
well understood \citep{SB14}.

Recently, \citet{Gent2021} sought to derive criteria for the appearance of
a small-scale dynamo in simulations of interstellar turbulence.
By not employing a shearing-box setup or stratification, they
focused only on the effects of the supernova-driven turbulence.
They confirm that,
below a critical physical resistivity (i.e., a sufficiently high $\Rm$),
a small-scale dynamo is easily excited by ISM turbulence, 
a result that appears to converge at resolutions below 1~pc.

\begin{figure}
\includegraphics[width=\textwidth]{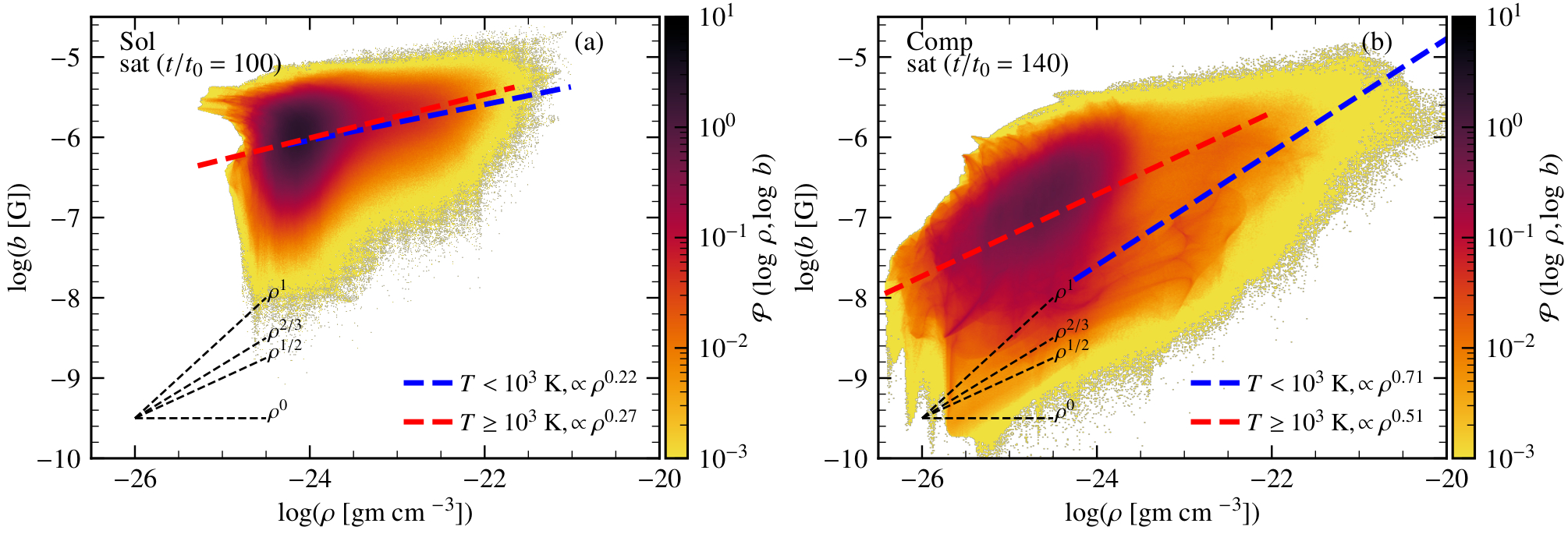}
\caption{
2D PDFs of magnetic field and density for (a) solenoidal and (b) compressive forcing.
The dashed black lines show various $b$--$\rho$ relations for simple gas compressions.
Especially the solenoidal shows very little similarity with any of the simple relations.
}\label{fig:Seta+Federrath22}
\end{figure}

\cite{Seta+Federrath22} have shown that the multiphase aspect of the
ISM tends to have a detrimental effect on the small-scale dynamo.
This is mostly because of the stronger Lorentz force in the
cold regions.
Their simulations show that with solenoidal forcing, the magnetic
field is mostly decoupled from the density behavior; see 
\Fig{fig:Seta+Federrath22}.
Simple compression along magnetic field lines ($\rho^{0}$),
perpendicular to to magnetic field lines ($b \propto \rho^{1/2}$
for cylindrical/filamentary geometry and $b \propto \rho^{1}$ for disc-like/slab geometry),
and spherical compression ($b \propto \rho^{2/3}$) hardly occur.
One might argue, however, that in the compressive case, the cold phase
shows a higher slope than the warm phase.

\subsection{Global isolated galaxy simulations}

The increasing efficiency and complexity of numerical codes and the
availability of resources have led to a number of works studying the
magnetic field evolution in global galaxy models.
In contrast to the shearing-box approach, such models more naturally allow
for the study of the large-scale dynamo.
However, the limited resolution is still problematic for simultaneously
capturing the small-scale dynamo, as outlined above.

\citet{Wang_Abel_2009} performed disk galaxy simulations including $\nG$
ordered seeds with a code similar to Enzo, using \citet{dedner_2002}
divergence-cleaning.
They found that the tiny seed was amplified to $\mu$G levels over 500~Myr.
They also noticed that the magnetic field in the cold gas saturated first.
Their setup did not include stellar feedback, so the amplification
process was driven by differential rotation only.

In a series of papers, \citet{Rieder_Teyssier_2016, Rieder_Teyssier_2017a,
Rieder_Teyssier_2017b} performed multi-component simulations of the
magnetic field evolution in dwarf and Milky-Way-like galaxies using the
RAMSES code and including supernova feedback.
Their setup includes dark matter and stars as collisionless particles,
coupled to an Adaptive Mesh Refinement (AMR) grid on which the MHD
equations are solved.
In the first paper of the series, the authors found signatures of
\begin{marginnote}
\entry{(Stellar) Feedback}{The energy and momentum deposited to the
ISM by stars through radiation, winds, and supernova explosions}
\end{marginnote}
small-scale dynamo amplification during intense feedback epochs, followed
by a large-scale dynamo at later, more quiescent evolution times.
In the second paper, they examined the saturation of the dynamo, which
occurs at only a small fraction of the turbulent kinetic energy.
They observed that, if the feedback efficiency is artificially lowered
after saturation, the turbulence decays and the galaxy settles in a thin
disk with an equipartition field.
\citet{Rieder_Teyssier_2017b} studied the magnetic
field evolution in a cosmological context.

Using a similar setup and separating the mean from the fluctuating
component using a median filter, \citet{Ntormousi+20} found large-scale
dynamo action in a model of a massive spiral.
However, their result was insensitive to the inclusion of supernova
feedback. 
Since supernova feedback is considered an important driver of small-scale
turbulence, this could mean that a small-scale dynamo was never captured
in their models.
This result is consistent with the \citet{Rieder_Teyssier_2016} results
for the quiescent phase of galaxy evolution.
However, as shown by \cite{Gent2021}, the limited resolution of the
simulation could be preventing the formation of a small-scale dynamo
action in this quiescent phase.

\citet{Pakmor2017} performed a suite of zoom-in cosmological
simulations that includes 30 galaxies using the AREPO code 
(called the Auriga simulations).
Similarly to \citet{Rieder_Teyssier_2017b}, they reported early
exponential growth of the magnetic field, saturating at redshift $z\simeq2-3$
at a few percent of the turbulent kinetic energy.
\citet{Steinwandel2019} also claimed significant small-scale dynamo
action in isolated galaxy models using an MHD version of the Gadget code.
However, these simulations rely on the divergence-cleaning scheme that could
suffer from the problems summarized in \Sec{NumericalApproaches}.

A very different approach was adopted by \citet{Rodriguesetal2019},
who modeled galactic magnetic fields by post-processing cosmological
simulation data.
Specifically, they inserted galaxy parameters such as shear rate and
turbulence into a parameter-fitting package that returns a suitable
dynamo solution \citep{Shukurov2019}.
Although this approach cannot capture the back-reaction of the magnetic
field on the gas, it can give estimates on the cosmological conditions
that favor mean-field dynamo action, which the authors find set in at
redshift $z<3$.

While each numerical approach presents certain limitations, the tentative
picture painted by global numerical simulations is that the small-scale
and large-scale dynamos co-exist during galaxy evolution.
Taking into account the large-scale gravitational collapse of the halo,
as well as internal galactic processes such as star formation, appears
to be fundamental for reconstructing the observed magnetic field evolution.

Currently, the problem of catastrophic quenching we discussed in
\Sec{Catastrophic_closed} remains unexplored in global galaxy models.
The reason is that physical resistivity and viscosity are usually not
included, so that there is no easy estimate of the $\Rm$ range probed
by each model.
Exploring the effects of the inevitable \emph{numerical} resistivity by
performing resolution studies might also be insufficient to approach the
physical solution, because the diffusion operator depends on resolution.
Small-scale helicity fluxes, which could in principle appear
self-consistently in these models, are not reported.
It would be interesting to see in upcoming studies how helicity fluxes
emerge (or not) from different subgrid models.

\section{INTERACTION WITH THE CGM}

It was already suggested in the previous sections that the galactic
environment must play an important role in the behavior of the dynamo,
because it defines the boundary conditions for its operation.
The immediate environment of a galaxy is 
its halo, which contains large amounts of diffuse gas, and is usually referred
to as the CGM. 

The CGM is a powerful probe of galaxy evolution processes, because it
contains traces of the cold ($T<10^5\K$) and hot ($T>10^5\K$) galactic
inflows, as well as the hot ($T\simeq10^6\K$), metal-enriched outflows
from feedback events \citep{Putman2012}.
The CGM also contains colder gas ($T<10^4\K$) that can co-exist with
these hotter phases for long periods of time.
This observation has led to theories involving magnetic fields and
cosmic rays in the dynamics of the CGM.

This cold gas was studied in the context of cosmological simulations by
\citet{Nelson2020}, who found small-scale cool ($T<10^4$ K) structures
in massive ($M\simeq10^{13}-10^{13.5}M_\odot$) galaxy halos.
In these simulations, the cloudlets are created by thermal instability,
seeded by tidally stripped gas from in-falling halos.
In many cases, these structures are dominated by magnetic pressure.
However, it is not clear whether these properties would persist at
higher resolution.
In spite of the dominant magnetic pressure, an otherwise identical
simulation of a massive halo with the magnetic field set to zero
showed essentially no change in the distribution and morphology of
these cloudlets.

\subsection{Magnetization of the galaxy by inflows}

The classical picture of gas accretion onto a galaxy halo predicts that the gas
should shock and heat up to high temperatures
\citep[e.g.][]{White_rees1978}.
However, cosmological simulations of galaxy formation
\citep[e.g.,][]{dekel2009} showed that a high fraction of the in-flowing
gas in high-redshift galaxies and present-day dwarfs is organized in cool
($T\simeq10^5\K$) streams.

RM observations suggest that this gas carries a $nG$-level magnetic
field \citep[e.g.,][]{Caretti2022}.
These estimates are compatible with the predictions of cosmological
magnetic field evolution models, which include detectable intergalactic
magnetic fields from the evolution of primordial seeds \citep[e.g.,][see
also \Sec{sec:cosmosims}]{vazza2015,Vazza2017}.
Then a primordial galaxy might receive a strong seed for its own
magnetic field.
However, at the time of this review, the possible effect of magnetized
inflowing gas on the galactic dynamo remains unexplored.
As we have seen in \Sec{sec:mean-field_models}, some models invoke a
large-scale dynamo in galactic halos to explain the observed X-shaped
magnetic fields therein.
If there is really large-scale dynamo action in the halo, it may be
predominantly of dipolar parity \citep{SS90}.
This can lead to an interaction and competition with the quadrupolar
magnetic field in the disk.
\cite{BDMSST92} found that, during certain time intervals, the RM
of these models shows a doubly peaked azimuthal variation,
which could be falsely interpreted as an indication of a bisymmetric
field structure.
The galactic halo may also act as a buffer for the dynamo in the disk
to dispose of excess magnetic helicity.

Galaxy mergers are a particular form of inflow, which can influence the
entire structure of the galaxy.
As the galaxies approach each other, their star formation rate is enhanced
and shocks form in their interstellar media.
Both shocks and small-scale flows associated with feedback from young stars
can strongly amplify the magnetic field.
This highly nonlinear type of interaction requires numerical modeling.

\begin{figure}
\includegraphics[width=\linewidth]{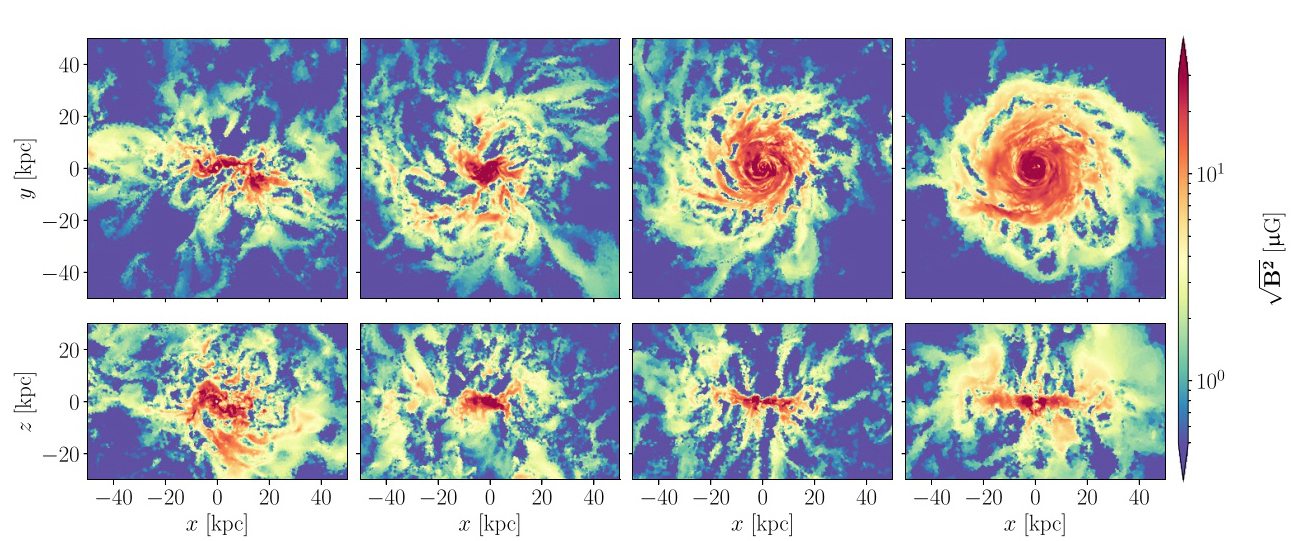}
\caption{
Adapted panel of Figure~3 from \citet{Whittingham2021}, showing a galaxy
re-arranging after a merger event.
The selected panel shows the magnetic field strength face-on and edge-on.
These simulations are part of the Auriga project \citep{grand2017}.
}\label{fig:magnetized_outflow}
\end{figure}

Most numerical models of galaxy mergers with magnetization so far were
done using Lagrangian codes, which have an obvious advantage in adapting their resolution in this setup.
The first numerical simulation of a galaxy merger with magnetic fields was
performed by \citet{Kotarba2010}, who modeled the Antennae galaxies using an MHD
version of the Gadget code that subtracts the Lorenz force associated
with magnetic divergence.
They confirmed the expected amplification of the magnetic field during
the galaxy encounters.
However, the amplification was also accompanied by a surge in numerical
magnetic divergence.
The subsequent cosmological merger models of \citet{Beck+2012} suffered
from the same issue.
\citet{Whittingham2021} showed more sophisticated merger models,
modeled in a cosmological context using the AREPO code.
They found a significant impact of the magnetic field on the morphology
of the remnant galaxy.
Specifically, a comparison between MHD and hydrodynamic models showed the
presence of extended disks and spiral structure in the magnetized mergers,
as opposed to compact remnants with a ring morphology in unmagnetized
mergers. 
\Fig{fig:magnetized_outflow}, from their
work, shows the evolution of the galaxy post-merger.

One exception to the Lagrangian models is the work of
\citet{Rodenbeck2016}, who performed a grid simulation of
a galaxy merger, using a simplified model without stellar
feedback or a collisionless component.
They found that the enhancement of the magnetic field is
particularly pronounced in the central regions of the galaxy.

\subsection{Magnetization of the CGM by outflows}

Galactic outflows are fundamental in any theory of galaxy evolution.
In starburst and dwarf galaxies they are powered by stellar feedback
\citep[e.g.,][]{Zhang2018}, while a few galaxies host AGN-powered winds
\citep{Fabian2012,Kormendy_Ho2013,Martin1998,Veilleux2005}.
Recent work has indicated that, in some cases, galactic winds can be
driven by cosmic ray (CR) pressure \citep{Hanasz2013,girichidis2016}.
The material carried by these outflows can magnetize the CGM.

In a recent study, \citet{vdv2021} performed direct galaxy evolution simulations
while resolving the magnetized CGM, as part of the Auriga project.
They found that, while the CGM remained a high-beta plasma, magnetic fields
can noticeably change its structure around galaxies, indirectly affecting
numerous galactic processes.
For instance, galactic outflows become more collimated, resulting in
less efficient mixing between enriched and un-enriched gas.
Outflow speeds are also reduced in the presence of magnetization,
which means that more metals remain in the halo with respect to the
un-magnetized situation.
The overall structure of the CGM is smoother, due to the additional
magnetic pressure.

\citet{Aramburo-Garcia2021} studied the magnetization of halos by AGN
and SN-driven outflows in the Illustris-TNG simulations.
They found that both types of outflows contributed to the creation
of over-magnetized bubbles, with the AGN-driven bubbles playing the
dominant role.

\subsection{The impact of the environment on the galactic dynamo}

The interaction of the galaxy with the CGM can have a crucial impact
on the development of a dynamo.
The loss of helicity flux through a galactic wind or fountain can
help avoid catastrophic quenching and sustain a dynamo for longer
(see \Sec{sec:helicityflux}), eventually reaching higher values of
the saturated field.
However, winds can also interfere with the dynamo itself if they are
acting within the dynamo-active region.
On the other hand, a strong magnetic field can suppress the galactic
outflow.

Whether we are considering inflows, outflows, or a galactic fountain, the
galaxy is always embedded in a current system that affects
the evolution of the dynamo.
This is a non-trivial complication of the effective boundary conditions
because the level of magnetization of the inflowing or outflowing gas
is unknown, and the extent of these flows can be much larger than the
virial radius of the galaxy.


\section{CONCLUSIONS}

After 70 years of inquiry into the possibility of dynamos in galaxies,
several important questions can now be answered.
In virtually all astrophysical settings, there is turbulence
and this turbulence is always magnetized because of small-scale dynamo
action.
Dynamos also work in decaying and otherwise nonstationary turbulence
and can produce equipartition-strength magnetic fields exponentially on
a turbulent turnover timescale.
This realization makes the question of {\em cosmological} seed magnetic fields
for galaxies and galaxy clusters almost obsolete, because they would always
be overpowered by small-scale dynamos that can operate very rapidly when
the scales are sufficiently small.
While primordial magnetic fields may still be present and interesting in
their own rights, the simple idea of them being wound up to explain the
bisymmetric spiral of nonaxisymmetric magnetic fields in some galaxies
is essentially ruled out.

While global numerical simulations are now beginning to show
the production of magnetic fields in galaxies, there remains a big
uncertainty regarding the question of what actually produces
large-scale dynamo action.
Is it really the $\alpha$ effect or some other mechanisms at play?
In this review, we have outlined several known mechanisms that could
produce large-scale magnetic fields, but this question remains a major
research topic in the years to come.
One reason behind this is that the problem of
catastrophic quenching is still not fully resolved.
Low-resolution simulations of relatively diffusive dynamos may have been
promising in terms of field strength and structure, but so far they have not
survived the test of higher resolution.
It remains important to continue to investigate this.
At the same time, it is important to open one's mind and think about
dynamos beyond just the immediate proximity of a galaxy.
The interaction with the CGM may be of crucial importance, and not all
of the different processes, which are important, may qualify as a dynamo.

\begin{summary}[SUMMARY POINTS]
\begin{enumerate}
\item Small-scale dynamos work in all turbulent astrophysical environments.
\item Simulations suggest that the efficiency of large-scale dynamos
decreases with increasing resolution, probably because magnetic helicity fluxes
are still inefficient.
\item The relevance of an $\alpha$ effect dynamo in galaxies remains unclear.
\item Modern numerical simulations of galactic magnetic fields tend to take the past evolution of and the interaction with the environment into account.
\end{enumerate}
\end{summary}

\begin{issues}[FUTURE ISSUES]
\begin{enumerate}
\item
The problem of catastrophic quenching remains relevant and will need to
be addressed in high-resolution models with realistic boundary conditions.
\item
Numerical codes of galaxy evolution should take special care of the
accurate treatment of magnetic fields, especially on the solenoidality
constraint, a problem that is accentuated through the subgrid modeling
of star formation and feedback.
\item The next generation of numerical models should make an effort to identify observables related to the galactic dynamo that go beyond the BSS/ASS signature in the RM of galaxies.
\item Future numerical models will have to quantify the importance of dynamo action in different stages of a galaxy's evolution. 
\end{enumerate}
\end{issues}

\section*{DISCLOSURE STATEMENT}

The authors are not aware of any affiliations, memberships, funding, or
financial holdings that might be perceived as affecting the objectivity
of this review.

\section*{ACKNOWLEDGMENTS}

We thank Eve Ostriker for a detailed review of the manuscript
and Rainer Beck, Oliver Gressel, Yik Ki (Jackie) Ma, Anvar Shukurov, and
Kandaswamy Subramanian for constructive comments on an earlier draft.
We acknowledge support for the Nordita program on ``Magnetic field
evolution in low density or strongly stratified plasmas'' in May 2022,
when part of this work was done.
EN also acknowledges Interstellar Institute's program "With Two Eyes"
and the Paris-Saclay University's Institut Pascal for hosting fruitful
discussions that nourished some ideas included in this work.
We acknowledge very useful discussions with Fabio Del Sordo, Frederick Gent, Sergio Martin-Alvarez, and Jennifer West.
EN acknowledges funding from the ERC Grant ”Interstellar” (Grant agreement 740120) and the Hellenic Foundation for Research and Innovation (Project number 224). 
This work was supported by the Swedish Research Council
(Vetenskapsr{\aa}det, 2019-04234).
Nordita is sponsored by Nordforsk.
We acknowledge the allocation of computing resources provided by the
Swedish National Allocations Committee at the Center for Parallel
Computers at the Royal Institute of Technology in Stockholm and
Link\"oping.

\bibliography{ref}{}
\bibliographystyle{ar-style2}
\end{document}

%% file: macros.tex
\newcommand{\EQ}{\begin{equation}}
\newcommand{\EN}{\end{equation}}
\newcommand{\EQA}{\begin{eqnarray}}
\newcommand{\ENA}{\end{eqnarray}}
\newcommand{\eq}[1]{\ref{#1}}
\newcommand{\EEq}[1]{Equation~\ref{#1}}
\newcommand{\Eq}[1]{Equation~\ref{#1}}

\newcommand{\Sec}[1]{Section~\ref{#1}}

\newcommand{\Fig}[1]{Figure~\ref{#1}}
\newcommand{\FFig}[1]{Figure~\ref{#1}}

\newcommand{\Tab}[1]{Table~\ref{#1}}

\newcommand{\bra}[1]{\langle #1\rangle}

\DeclareMathAlphabet\mathbfcal{OMS}{cmsy}{b}{n}
{}
{}
\newcommand{\FFFF}{\mathbfcal{F}}

\newcommand{\meanFFFFm}{\overline{\mathbfcal{F}}_{\rm m}}
\newcommand{\meanFFFFf}{\overline{\mathbfcal{F}}_{\rm f}}
\newcommand{\meanemf}{\overline{\cal E} {}}

{}
{}
\newcommand{\meanEMF}{\overline{\mathbfcal{E}}}
{}
{}
{}
{}
{}
{}
{}
{}
{}
{}
{}
{}
{}
{}
\newcommand{\meanUU}{\overline{\bm{U}}}

{}

\newcommand{\meanB}{\overline{B}}

\newcommand{\meanU}{\overline{U}}

\newcommand{\meanJ}{\overline{J}}

{}

{}
{}
{}

\newcommand{\zzz}{\hat{\mbox{\boldmath $z$}} {}}

\newcommand{\meanAA}{{\overline{\bm{A}}}}
\newcommand{\meanBB}{{\overline{\bm{B}}}}
\newcommand{\meanJJ}{{\overline{\bm{J}}}}

\newcommand{\xx}{\bm{x}}

\newcommand{\aaaa}{\bm{a}}
\newcommand{\jj}{\bm{j}}

\newcommand{\grav}{\bm{g}}
\newcommand{\bb}{\bm{b}}
\newcommand{\BB}{\bm{B}}

\newcommand{\EE}{\bm{E}}
\newcommand{\JJ}{\bm{J}}
\newcommand{\oo}{\bm{\omega}}

\newcommand{\AAA}{\bm{A}}

\newcommand{\UU}{\bm{U}}

\newcommand{\uu}{\bm{u}}

\newcommand{\nab}{{\bm{\nabla}}}

\newcommand{\OO}{\bm{\Omega}}

\newcommand{\pom}{\mbox{\boldmath $\varpi$} {}}

\newcommand{\ggamma}{\mbox{\boldmath $\gamma$} {}}

\newcommand{\SSSS}{\mbox{\boldmath ${\sf S}$} {}}

\newcommand{\ii}{{\rm i}}

\newcommand{\DD}{{\rm D} {}}

\newcommand{\dd}{{\rm d} {}}

\newcommand{\const}{{\rm const}  {}}

\def\degr{\hbox{$^\circ$}}

\def\Pm{\mbox{\rm Pr}_{\rm M}}
\def\Rm{\mbox{\rm Re}_{\rm M}}

\def\Rmc{\mbox{\rm Re}_{\rm M}^{\rm crit}}

\def\Rey{\mbox{\rm Re}}
\def\Imag{\mbox{\rm Im}}

\def\EM{E_{\rm M}}

\def\kf{k_{\rm f}}

\def\EM{E_{\rm M}}

\def\epsK{\epsilon_{\rm K}}

\def\urms{u_{\rm rms}}

\def\orms{\omega_{\rm rms}}

\def\etat{\eta_{\rm t}}

\def\etaT{\eta_{\rm T}}

\def\Beq{B_{\rm eq}}

\newcommand{\G}{\,{\rm G}}

\newcommand{\K}{\,{\rm K}}

\newcommand{\s}{\,{\rm s}}

\newcommand{\uG}{\,\mu{\rm G}}
\newcommand{\nG}{\,{\rm nG}}

\newcommand{\cm}{\,{\rm cm}}

\newcommand{\m}{\,{\rm m}}
\newcommand{\km}{\,{\rm km}}

\newcommand{\pc}{\,{\rm pc}}
\newcommand{\kpc}{\,{\rm kpc}}

\newcommand{\yr}{\,{\rm yr}}

\newcommand{\Gyr}{\,{\rm Gyr}}

